\documentstyle[pra,epsf,cite,eqsecnum,aps,multicol]{revtex}

\renewcommand{\narrowtext}{\begin{multicols}{2} \global\columnwidth20.5pc}
\renewcommand{\widetext}{\end{multicols} \global\columnwidth42.5pc}
\def\lfrac#1#2{{{\scriptstyle #1}\over {\scriptstyle #2}}}

\draft

\begin{document}

\title{Electron Self Energy for the K and L Shell at Low Nuclear Charge}

\author{Ulrich D.\ Jentschura,$^{1,2,}$\cite{InternetUJ}, 
Peter J.\ Mohr,$^{1,}$\cite{InternetPJM},
and Gerhard Soff$^{\,2,}$\cite{InternetGS}}

\address{$^1$National Institute of Standards and Technology, 
Mail Stop 8401, Gaithersburg, MD 20899-8401, USA \\
$^2$Institut f\"{u}r Theoretische Physik, TU Dresden,
Mommsenstra\ss e 13, 01062 Dresden, Germany}

\maketitle

\begin{abstract}
A nonperturbative numerical evaluation of the one-photon electron self
energy for the K- and L-shell states of hydrogenlike ions with nuclear
charge numbers $Z=1$ to 5 is described. Our calculation for the $1{\rm
S}_{1/2}$ state has a numerical uncertainty of 0.8~Hz in atomic
hydrogen, and for the L-shell states ($2{\rm S}_{1/2}$, $2{\rm
P}_{1/2}$, and $2{\rm P}_{3/2}$) the numerical uncertainty is 1.0~Hz.
The method of evaluation for the ground state and for the excited
states is described in detail. The numerical results are compared to
results based on known terms in the expansion of the self energy in
powers of $Z\alpha$.
\end{abstract}


\narrowtext

\tableofcontents

\typeout{Section 1}
%
%
\section{Introduction}

The nonperturbative numerical evaluation of radiative corrections to
bound-state energy levels is interesting for two reasons. First, the
recent dramatic increase in the accuracy of experiments that measure
the transition frequencies in hydrogen and
deuterium~\cite{BeEtAl1997,UdEtAl1997,NiEtAl2000} necessitates a numerical
evaluation (nonperturbative in the binding Coulomb field) of the
radiative corrections to the spectrum of atomic systems with low
nuclear charge $Z$. Second, the numerical calculation serves as an
independent test of analytic evaluations which are based on an
expansion in the binding field with an expansion parameter $Z\alpha$.

In order to address both issues, a high-precision numerical evaluation
of the self energy of an electron in the ground state in hydrogenlike
ions has been performed~\cite{JeMoSo1999,Je1999}. The approach
outlined in~\cite{JeMoSo1999} is generalized here to the L shell, and
numerical results are obtained for the ($n=2$) states $2{\rm
S}_{1/2}$, $2{\rm P}_{1/2}$ and $2{\rm P}_{3/2}$. Results are provided
for atomic hydrogen, ${\rm He}^{+}$, ${\rm Li}^{2+}$, ${\rm Be}^{3+}$,
and ${\rm B}^{4+}$.

It has been pointed out in~\cite{JeMoSo1999,Je1999} that the
nonperturbative effects (in $Z\alpha$) can be large even for
low nuclear charge and exceed the current experimental accuracy for 
atomic transitions. For example, the difference between the sum of the
analytically evaluated terms up to the order of $\alpha \,
(Z\alpha)^6$ and the final numerical result for the ground state is
roughly 27~kHz for atomic hydrogen and about 3200~kHz for ${\rm
He}^{+}$.  For the $2{\rm S}$ state the difference is $3.5$~kHz for
atomic hydrogen and 412~kHz for ${\rm He}^{+}$. The large difference
between the result obtained by an expansion in $Z\alpha$ persists even
after the inclusion of a result recently obtained in~\cite{Ka1997} for
the logarithmic term of order $\alpha\,(Z\alpha)^7\,\ln(Z\alpha)^{-2}$. 
For the ground state, the difference between the all-order numerical
result and the sum of the perturbative terms is still 13~kHz for
atomic hydrogen and 1600~kHz for ${\rm He}^{+}$. For the $2\,{\rm S}$
state, the difference amounts to $1.6$~kHz for atomic hydrogen and to
213~kHz for ${\rm He}^{+}$.

These figures should be compared to the current experimental
precision. The most accurately measured transition to date is the
1S--2S frequency in hydrogen; it has been measured with a
relative uncertainty of $1.8$ parts in $10^{14}$ or $46~{\rm
Hz}$~\cite{NiEtAl2000}. This experimental progress is due in part to
the use of frequency chains that bridge the range between optical
frequencies and the microwave cesium time standard. The uncertainty of
the measurement is likely to be reduced by an order of magnitude in
the near future~\cite{NiEtAl2000,HaPr2000}.  With trapped hydrogen atoms, it
should be feasible to observe the 1S--2S frequency with an
experimental linewidth that approaches the $1.3\,{\rm Hz}$ natural
width of the 2S level \cite{CeEtAl1996,KiEtAl1998}.

The perturbation series in $Z\alpha$ is slowly convergent. The
all-order numerical calculation presented in this paper essentially
eliminates the uncertainty from unevaluated higher-order analytic
terms, and we obtain results for the self-energy remainder function
$G_{\rm SE}$ with a precision of roughly $0.8 \times Z^4 \, {\rm Hz}$
for the ground state of atomic hydrogen and $1.0 \times Z^4 \, {\rm
Hz}$ for the $2{\rm S}$ state.

In the evaluation, we take advantage of resummation and convergence
acceleration techniques. The resummation techniques provide an
efficient method of 
evaluation of the Dirac-Coulomb Green function to a relative
uncertainty of $10^{-24}$ over a wide parameter range~\cite{Je1999}. The
convergence acceleration techniques remove the principal numerical
difficulties associated with the singularity of the relativistic
propagators for nearly equal radial arguments~\cite{JeMoSoWe1999}.

The one-photon self energy treated in the current investigation is
about two orders of magnitude larger than the other contributions
to the Lamb shift in atomic hydrogen.
A comprehensive review of the various
contributions to the Lamb shift in hydrogenlike atoms in the full
range of nuclear charge numbers $Z=1$--$110$ has been given
in~\cite{JoSo1985,MoPlSo1998,Mo1996,EiGrSh2000}.

This paper is organized as follows. The method of evaluation is
discussed in Sec.~\ref{MethodOfEvaluation}. The calculation is divided
into a low-energy part and a high-energy contribution.  The low-energy
part is treated in Sec.~\ref{LowEnergyPart}, and the high-energy part
is discussed in Sec.~\ref{HighEnergyPart}. Numerical results are
compiled in Sec.~\ref{ComparisonAnalytic}. Also in
Sec.~\ref{ComparisonAnalytic}, we compare numerical and analytic
results for the Lamb shift in the region of low nuclear charge
numbers. Of special importance is the consistency
check with available analytic results~\cite{Pa1993,JePa1996} for
higher-order binding corrections to the Lamb shift.  We make
concluding remarks in Sec.~\ref{Conclusion}.

\typeout{Section 2}
%
%
\section{Method of Evaluation}
\label{MethodOfEvaluation}

\typeout{Subsection 2A}
%
%
\subsection{Status of Analytic Calculations}
\label{StatusAnalytic}

The (real part of the) energy shift $\Delta E_{\rm SE}$ due to the
electron self-energy radiative correction is usually written as
\begin{equation}
\label{ESEasF}
\Delta E_{\rm SE} = \frac{\alpha}{\pi} \, \frac{(Z \alpha)^4}{n^3} \, 
F(nl_j,Z\alpha) \, m_{\rm e} \, c^2
\end{equation}
where $F$ is a dimensionless quantity. In the following, the natural
unit system with $\hbar = c = m_{\rm e} = 1$ and $e^2 = 4\pi\alpha$ is
employed.  Note that $F(nl_j,Z\alpha)$ is a dimensionless function
which depends for a given atomic state with quantum numbers $n$, $l$
and $j$ on only one argument (the coupling $Z\alpha$).  For excited
states, the (nonvanishing) imaginary part of the self energy
is proportional to the (spontaneous) decay width of the state. We will
denote here the {\em real} part of the self energy by $\Delta E_{\rm
SE}$, exclusively.  The semi-analytic expansion of $F(nl_j,Z\alpha)$
about $Z\alpha=0$ for a general atomic state with quantum numbers $n$,
$l$ and $j$ gives rise to the semi-analytic expansion,
\begin{eqnarray}
\label{defFLO}
\lefteqn{F(nl_j,Z\alpha) =
A_{41}(nl_j) \, \ln(Z \alpha)^{-2}} \nonumber\\[2ex]
& & \;\; + A_{40}(nl_j) +
(Z \alpha) \, A_{50}(nl_j) \nonumber\\[2ex]
& & \;\; + \,
(Z \alpha)^2 \,
\left[A_{62}(nl_j) \, \ln^2(Z \alpha)^{-2} \right.
\nonumber\\[2ex]
& & \;\;\;\; \left. + A_{61}(nl_j) \,\ln(Z \alpha)^{-2} +
G_{\rm SE}(nl_j,Z\alpha) \right]\,.
\end{eqnarray}
For particular states, some of the coefficients may vanish.  Notably,
this is the case for P states, which are less singular than S states
at the origin [see Eq.~(\ref{defFLOnPj}) below].  For the $n{\rm
S}_{1/2}$ state ($l=0$, $j=1/2$), none of the terms in
Eq.~(\ref{defFLO}) vanishes, and we have,
\begin{eqnarray}
\label{defFLOnS}
\lefteqn{F(n{\rm S}_{1/2},Z\alpha) = 
A_{41}(n{\rm S}_{1/2}) \, \ln(Z \alpha)^{-2}} \nonumber\\[2ex]
& & \;\; + A_{40}(n{\rm S}_{1/2}) +
(Z \alpha) \, A_{50}(n{\rm S}_{1/2}) \nonumber\\[2ex]
& & \;\; + \, 
(Z \alpha)^2 \, 
\left[A_{62}(n{\rm S}_{1/2}) \, \ln^2(Z \alpha)^{-2} \right. 
\nonumber\\[2ex]
& & \;\;\;\; \left. + A_{61}(n{\rm S}_{1/2}) \,\ln(Z \alpha)^{-2} + 
G_{\rm SE}(n{\rm S}_{1/2},Z\alpha) \right]\,.
\end{eqnarray}
The $A$ coefficients have two indices, the first of which denotes the
power of $Z\alpha$ [including those powers implicitly contained in
Eq.~(\ref{ESEasF})], while the second index denotes the power of the
logarithm $\ln(Z \alpha)^{-2}$.  For P states, the coefficients
$A_{41}$, $A_{50}$ and $A_{62}$ vanish, and we have
\begin{eqnarray}
\label{defFLOnPj}
\lefteqn{F(n{\rm P}_j,Z\alpha) =
A_{40}(n{\rm P}_j)} \nonumber\\[2ex] 
& & \;\; + (Z \alpha)^2 \,
\left[A_{61}(n{\rm P}_j) \, \ln(Z \alpha)^{-2} +
G_{\rm SE}(n{\rm P}_j,Z\alpha) \right]\,.
\end{eqnarray}
For S states, the self-energy remainder function $G_{\rm SE}$ can be
expanded semi-analytically as
\begin{eqnarray}
\label{DefinitionOfA601S}
\lefteqn{G_{\rm SE}(n{\rm S}_{1/2},Z\alpha) = A_{60}(n{\rm S}_{1/2})}
\nonumber\\[2ex]
& & \quad\quad 
+ (Z\alpha)\,\left[A_{71}(n{\rm S}_{1/2})\,\ln(Z \alpha)^{-2}\right.
 \nonumber\\[2ex]
& & \quad\quad\quad\quad\quad\quad \left. 
+ A_{70}(n{\rm S}_{1/2}) + {\rm o}(Z\alpha)\right]
\end{eqnarray}
(for the ``order'' symbols o and O we follow the usual convention, see
e.g.~\cite{WhWa1944,Er1987}). For P states, the semi-analytic expansion of
$G_{\rm SE}$ reads
\begin{eqnarray}
\label{A60nP}
\lefteqn{G_{\rm SE}(n{\rm P}_j,Z\alpha) = A_{60}(n{\rm P}_j)}
\nonumber\\[2ex]
& & \quad + (Z\alpha)\,\left[A_{70}(n{\rm P}_j) 
  + {\rm o}(Z\alpha)\right]\,.
\end{eqnarray}
The fact that $A_{71}(n{\rm P}_j)$ vanishes has been pointed out
in~\cite{Ka1997}.  We list below the analytic coefficients and the
Bethe logarithms relevant to the atomic states under
investigation. For the ground state, the coefficients $A_{41}$ and
$A_{40}$ were obtained
in~\cite{Be1947,Fe1948,Fe1949,FrWe1949,KrLa1949,Sc1949,FuMiTo1949},
the correction term $A_{50}$ was found
in~\cite{Ba1951,KaKlSc1952,BaBeFe1953}, and the higher-order binding
corrections $A_{62}$ and $A_{61}$ were evaluated
in~\cite{FrYe1958,FrYe1960,La1960,La1961a,La1961b,
ErYe1965a,ErYe1965b,Er1971,Sa1981,Pa1993}. The results are,
\begin{eqnarray}
\label{coeffs1S12}
A_{41}(1{\rm S}_{1/2}) & = & \frac{4}{3}\,, \nonumber\\[1ex]
A_{40}(1{\rm S}_{1/2}) & = & 
  \frac{10}{9} - \frac{4}{3} \, \ln k_0(1{\rm S})\,, 
\nonumber\\[1ex]
A_{50}(1{\rm S}_{1/2}) & = & 
4\pi\,\left[\frac{139}{128} - \frac{1}{2}\,\ln 2\right]\,,
\nonumber\\[1ex]
A_{62}(1{\rm S}_{1/2}) & = & -1\,, \nonumber\\[1ex]
A_{61}(1{\rm S}_{1/2}) & = & \frac{28}{3} \, \ln 2 - \frac{21}{20}\,.
\end{eqnarray}
The Bethe logarithm $\ln k_0(1{\rm S})$ has been evaluated 
in~\cite{KlMa1973}
and \cite{BeBrSt1950,Ha1956,ScTi1959,Li1968,Hu1969} as
\begin{equation}
\label{BetheLog1S}
\ln k_0(1{\rm S}) = 2.984~128~555~8(3).
\end{equation}
For the 2S state, we have
\begin{eqnarray}
\label{coeffs2S12}
A_{41}(2{\rm S}_{1/2}) & = & \frac{4}{3}\,, \nonumber\\[1ex]
A_{40}(2{\rm S}_{1/2}) & = &
  \frac{10}{9} - \frac{4}{3} \, \ln k_0(2{\rm S})\,,
\nonumber\\[1ex]
A_{50}(2{\rm S}_{1/2}) & = &
4\pi\,\left[\frac{139}{128} - \frac{1}{2}\,\ln 2\right]\,,
\nonumber\\[1ex]
A_{62}(2{\rm S}_{1/2}) & = & -1\,, \nonumber\\[1ex]
A_{61}(2{\rm S}_{1/2}) & = & \frac{16}{3} \, \ln 2 + \frac{67}{30}\,.
\end{eqnarray}
The Bethe logarithm $\ln k_0(2{\rm S})$ has been evaluated
(see~\cite{KlMa1973,BeBrSt1950,Ha1956,ScTi1959,Li1968,Hu1969},
the results exhibit varying accuracy) as
\begin{equation}
\label{BetheLog2S}
\ln k_0(2{\rm S}) = 2.811~769~893(3).
\end{equation}
It might be worth noting that
the value for $\ln k_0(2{\rm S})$ given in~\cite{Er1977} evidently
contains a typographical error. Our independent re-evaluation
confirms the result given in Eq.~(\ref{BetheLog2S}), which was
originally obtained in~\cite{KlMa1973} to the required precision. For
the $2{\rm P}_{1/2}$ state we have
\begin{eqnarray}
\label{coeffs2P12}
A_{40}(2{\rm P}_{1/2}) & = &
  -\frac{1}{6} - \frac{4}{3} \, \ln k_0(2{\rm P})\,,
\nonumber\\[1ex]
A_{61}(2{\rm P}_{1/2}) & = & \frac{103}{108}\,.
\end{eqnarray}
Note that a general analytic result for the logarithmic correction
$A_{61}$ as a function of the bound state quantum numbers $n$, $l$ and
$j$ can be inferred from Eq.~(4.4a) of~\cite{ErYe1965a,ErYe1965b} upon
subtraction of the vacuum polarization contribution implicitly
contained in the quoted equation. The Bethe logarithm for the 2P
states reads~\cite{KlMa1973,DrSw1990}
\begin{equation}
\label{BetheLog2P}
\ln k_0(2{\rm P}) = -0.030~016~708~9(3)\,.
\end{equation}
Because the Bethe logarithm is an inherently nonrelativistic
quantity, it is spin-independent and therefore independent
of the total angular momentum $j$ for a given orbital angular momentum
$l$. For the $2{\rm
P}_{3/2}$ state the analytic coefficients are
\begin{eqnarray}
\label{coeffs2P32}
A_{40}(2{\rm P}_{3/2}) & = &
  \frac{1}{12} - \frac{4}{3} \, \ln k_0(2{\rm P})\,,
\nonumber\\[1ex]
A_{61}(2{\rm P}_{3/2}) & = & \frac{29}{90}\,.
\end{eqnarray}
We now consider the limit of the function $G_{\rm SE}(Z\alpha)$ as
$Z\alpha \to 0$.  The higher-order terms in the potential expansion
(see Fig.~\ref{ExactVExpansion} below) and relativistic corrections to
the wavefunction both generate terms of higher order in $Z\alpha$
which are manifest in Eq.~(\ref{defFLO}) in the form 
of the nonvanishing function $G_{\rm SE}(Z\alpha)$
which summarizes the effects of the relativistic corrections
to the bound electron wave function
and of higher-order terms in the potential expansion. For very soft virtual
photons, the potential expansion fails and generates an infrared
divergence which is cut off by the atomic momentum scale,
$Z\alpha$. This cut-off for the {\em infrared} divergence is one of
the mechanisms which lead to the logarithmic terms in
Eq.~(\ref{defFLO}). Some of the nonlogarithmic terms of relative order
$(Z\alpha)^2$ in Eq.~(\ref{defFLO}) are generated by the relativistic
corrections to the wave function. The function $G_{\rm SE}$ does not
vanish, but approaches a constant in the limit $Z\alpha\to 0$. This
constant can be determined by analytic or semi-analytic calculations;
it is referred to as the $A_{60}$ coefficient, i.e.
\begin{equation}
A_{60}(nl_j) = G_{\rm SE}(nl_j,0)\,.
\end{equation}
The evaluation of the coefficient $A_{60}(1{\rm S}_{1/2})$ has been
historically problematic~\cite{ErYe1965a,ErYe1965b,
Er1971,Sa1981,Pa1993}.
For the 2S state, there is currently only one precise analytic result
available,
\begin{equation}
\label{DiffA602S12}
\begin{array}{l@{\;\;\;=\;\;\;}l@{\;\;\;}l}
A_{60}(2{\rm S}_{1/2}) & -31.840\,47(1) &
  \mbox{\cite{Pa1993}\,.}
\end{array}\nonumber
\end{equation}
For the $2{\rm P}_{1/2}$ state, the analytically obtained result is
\begin{equation}
\label{A602P12}
\begin{array}{l@{\;\;\;=\;\;\;}l@{\;\;\;}l}
A_{60}(2{\rm P}_{1/2}) & -0.998\,91(1) &
  \mbox{\cite{JePa1996}\,,} 
\end{array}
\end{equation}
and for the $2{\rm P}_{3/2}$ state, we have 
\begin{equation}
\label{A602P32}
\begin{array}{l@{\;\;\;=\;\;\;}l@{\;\;\;}l}
A_{60}(2{\rm P}_{3/2}) & -0.503\,37(1) &
  \mbox{\cite{JePa1996}\,.}
\end{array}
\end{equation}
The analytic evaluations essentially rely on an expansion of the
relativistic Dirac-Coulomb propagator in powers of the binding field,
i.e.~in powers of Coulomb interactions of the electron with the
nucleus.  In numerical evaluations, the binding field is treated
nonperturbatively, and no expansion is performed.

\typeout{Subsection 2B}
%
%
\subsection{Formulation of the Numerical Problem}
\label{FormulationNumerical}

Numerical cancellations are severe for small nuclear charges.  In
order to understand the origin of the numerical cancellations it is
necessary to consider the renormalization of the self energy.  The
renormalization procedure postulates that the self energy is
essentially the effect on the bound electron due to the self
interaction with its own radiation field, minus the same effect on a
free electron which is absorbed in the mass of the electron and
therefore not observable. 
The self energy of the bound electron is the residual
effect obtained after the subtraction of two large quantities. Terms
associated with renormalization counterterms are of order $1$ in the
$Z\alpha$-expansion, whereas the residual effect is of order
$(Z\alpha)^4$ [see Eq.~(\ref{ESEasF})]. This corresponds to a loss of
roughly $9$ significant digits at $Z=1$. Consequently, even the
precise evaluation of the one-photon self energy in a Coulomb field
presented in \cite{Mo1992} extends only down to $Z = 5$. Among the
self-energy corrections in one-loop and higher-loop order, numerical
cancellations in absolute terms are most severe for the {\em one}-loop
problem because of the large size of the effect of the one-loop
self-energy correction on the spectrum.

%
%
\begin{figure}[htb]
\begin{center}
\begin{minipage}{8.6cm}
\centerline{\mbox{\epsfysize=7.0cm\epsffile{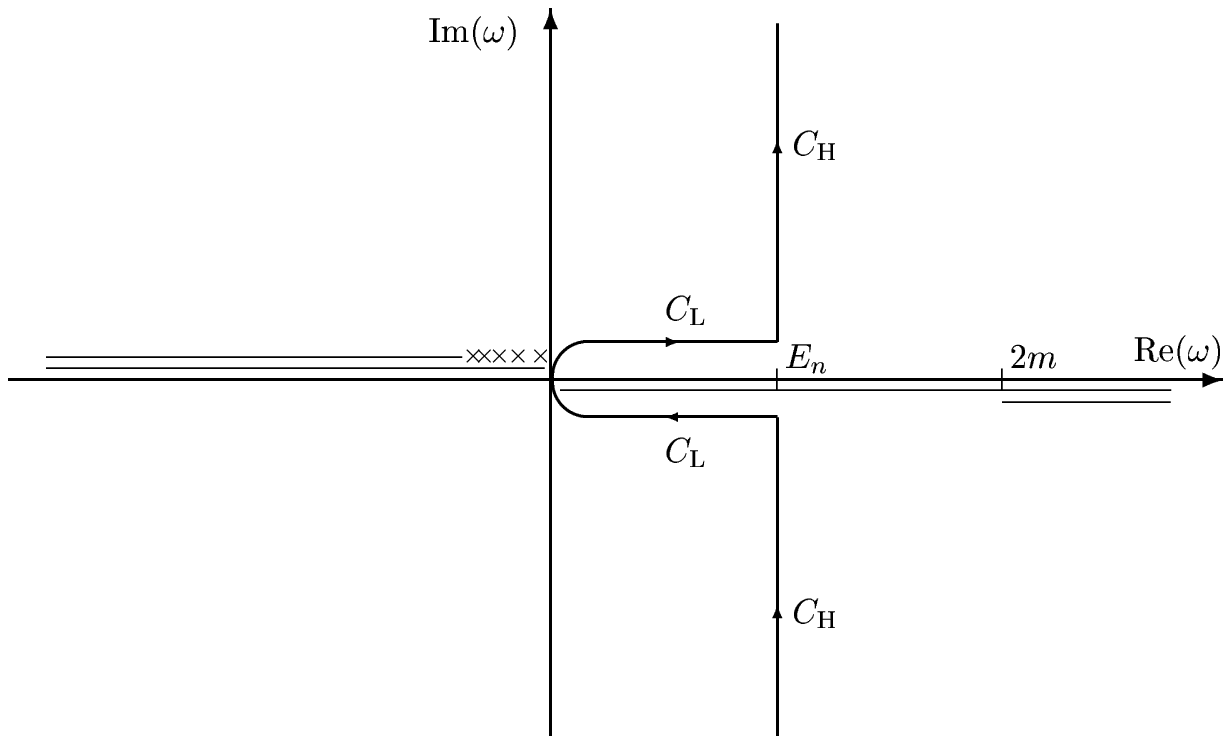}}}
\caption{\label{IntegrationContour}
Integration contour ${\cal C}$ for the integration over the energy
$\omega = E_n - z$ of the virtual photon. The contour ${\cal C}$
consists of the low-energy contour $C_{\rm L}$ and the high-energy
contour $C_{\rm H}$. Lines shown displaced
directly below and above the real axis
denote branch cuts from the photon and electron propagator. Crosses
denote poles originating from the discrete spectrum of the electron
propagator.  The contour used in this work corresponds to the one used
in~{\protect\cite{Mo1974a}}.}
\end{minipage}
\end{center}
\end{figure}

For our high-precision numerical evaluation, we start from the
regularized and renormalized expression for the one-loop self energy
of a bound electron,
\begin{eqnarray}
\label{deltaESEM}
\lefteqn{\Delta E_{\rm SE} = \lim_{\Lambda\to\infty}
\left\{ {\rm i}\, e^2 \, {\rm Re} \,
\int_{C_{\rm F}} \! \frac{d\omega}{2 \pi} 
\int \! \! \frac{d^3 \bbox{k}}{(2 \pi)^3} \,
D_{\mu\nu}(k^2,\Lambda) \right. } \nonumber\\[1ex]
& & \left. \times \left< \bar{\psi} \left| \gamma^{\mu} \,
  \frac{1}{\not{\! p} - \not{\! k} - 1 - \gamma^0 V }
    \gamma^{\nu} \, \right| \psi \right>  - \Delta m \right\} 
\nonumber\\[2ex]
&=& \lim_{\Lambda\to\infty}
\bigg\{ {\rm -i}\, e^2 \, {\rm Re} \,
\int_{\cal C} \! \frac{d\omega}{2 \pi}
\int \! \! \frac{d^3 \bbox{k}}{(2 \pi)^3} \,
D_{\mu\nu}(k^2,\Lambda)  \nonumber\\[1ex]
& & \!\!\!\!\!\!\!\!\! \, \times \left< \psi \left| \, \alpha^{\mu} \,
{\rm e}^{{\rm i}\bbox k\cdot \bbox x} \,
G(E_n - \omega) \, \alpha^{\nu} \, 
{\rm e}^{-{\rm i}\bbox k\cdot \bbox x} \,
\right| \psi \right>  - \Delta m \bigg\}\,,
\end{eqnarray}
where $G$ denotes the Dirac-Coulomb propagator,
\begin{equation}
\label{DefinitionOfG}
G(z) \; = \; \frac{1}{\bbox{\alpha}\cdot \bbox{p}
  + \beta + V - z }\,,
\end{equation}
and $\Delta m$ is the $\Lambda$-dependent 
(cutoff-dependent) one-loop mass-counter term,
\begin{equation}
\label{MassCounterTerm}
\Delta m = \frac{\alpha}{\pi} \,
\left(\frac{3}{4}\,\ln \Lambda^2 + \frac{3}{8} \right) \,
\langle \beta \rangle\,.
\end{equation}
The photon propagator $D_{\mu\nu}(k^2,\Lambda)$ in
Eq.~(\ref{deltaESEM}) in Feynman gauge reads
\begin{equation}
\label{PhotonPropagator}
D_{\mu\nu}(k^2,\Lambda) = 
  - \left(\frac{g_{\mu\nu}}{k^2 + {\rm i}\,\epsilon} - 
          \frac{g_{\mu\nu}}{k^2 - \Lambda^2 + {\rm i}\,\epsilon}\right)\,.
\end{equation}
The contour $C_{\rm F}$ in Eq.~(\ref{deltaESEM}) is the Feynman
contour, whereas the contour $\cal C$ is depicted in
Fig.~\ref{IntegrationContour}.  The contour $\cal C$ is employed for
the $\omega$-integration in the current evaluation [see the last line
of Eq.~(\ref{deltaESEM})]. The energy variable $z$ in
Eq.~(\ref{DefinitionOfG}) therefore assumes the value
\begin{equation} 
\label{DefOfZ}
z = E_n - \omega\,,
\end{equation}
where $E_n$ is the Dirac energy of the atomic state, and $\omega$
denotes the complex-valued energy of the virtual photon. It is
understood that the limit $\Lambda~\to~\infty$ is taken {\em after}
all integrals in Eq.~(\ref{deltaESEM}) are evaluated.

%
%
\begin{figure}[thb]
\begin{center}
\begin{minipage}{8.6cm}
\centerline{\mbox{\epsfysize=4.0cm\epsffile{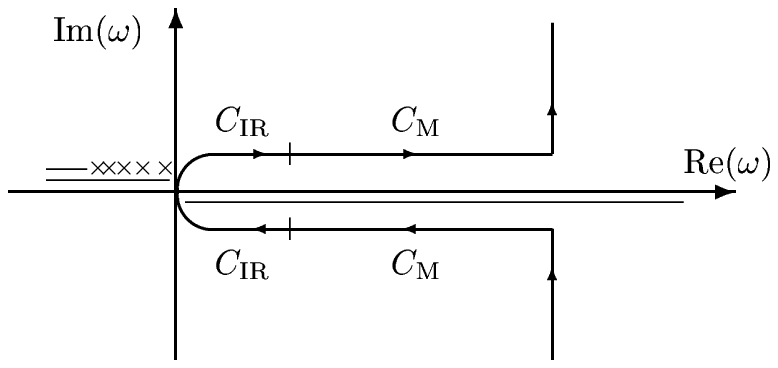}}}
\caption{\label{SeparationLow} 
Separation of the low-energy contour $C_{\rm L}$ into the infrared
part $C_{\rm IR}$ and the middle-energy part $C_{\rm M}$.  As in
Fig.~\ref{IntegrationContour}, the lines directly above and below the
real axis denote branch cuts from the photon and electron
propagator. Strictly speaking, the figure is valid only for the ground
state. For excited states, some of the crosses, which denote poles
originating from the discrete spectrum of the electron propagator, are
positioned to the right of the line ${\rm Re} \, \omega = 0$.  These
poles are subtracted in the numerical evaluation.}
\end{minipage}
\end{center}
\end{figure}

The integration contour for the complex-valued energy of the virtual
photon $\omega$ in this calculation is the contour $\cal C$ employed
in \cite{Mo1974a,Mo1974b,Mo1982,Mo1992} and depicted in
Fig.~\ref{IntegrationContour}.  The integrations along the low-energy
contour $C_{\rm L}$ and the high-energy contour $C_{\rm H}$ in
Fig.~\ref{IntegrationContour} give rise to the low- and the
high-energy contributions $\Delta E_{\rm L}$ and $\Delta E_{\rm H}$ to
the self energy, respectively. Here, we employ a further separation of
the low-energy integration contour $C_{\rm L}$ into an infrared
contour $C_{\rm IR}$ and a middle-energy contour $C_{\rm M}$ shown in
Fig.~\ref{SeparationLow}.  This separation gives rise to a separation
of the low-energy part $\Delta E_{\rm L}$ into the infrared part
$\Delta E_{\rm IR}$ and the middle-energy part $\Delta E_{\rm M}$,
\begin{equation}
\Delta E_{\rm L}=\Delta E_{\rm IR}+\Delta E_{\rm M}\,.
\end{equation}
For the low-$Z$ systems discussed here, all complications which arise
for excited states due to the decay into the ground state are
relevant only for the infrared part.  Except for the further
separation into the infrared and the middle-energy part, the same
basic formulation of the self-energy problem as in~\cite{Mo1974a} is
used.  This leads to the following separation:
\[
\begin{array}{ll}
\omega \in (0,\lfrac{1}{10}\,E_n) \pm {\rm i}\,\delta & ;  
\mbox{infrared part $\Delta E_{\rm IR}$,} \\[1ex]
\omega \in (\lfrac{1}{10}\,E_n, E_n) \pm {\rm i}\,\delta & ; 
\mbox{middle-energy part $\Delta E_{\rm M}$,} \\[1ex]
\omega \in E_n + {\rm i}\,(-\infty,+\infty) & ;  
\mbox{high-energy part $\Delta E_{\rm H}$.}
\end{array}
\]
Integration along these contours gives rise to the infrared, the
middle-energy, and the high-energy contributions to the energy
shift. For all of these contributions, lower-order terms are
subtracted in order to obtain the contribution to the self energy
of order $(Z\alpha)^4$. We obtain for the infrared part,
\begin{eqnarray}
\label{DefinitionOfFIR}
\Delta E_{\rm IR} &=& \frac{\alpha}{\pi} \,
\left[
\frac{21}{200} \langle \beta \rangle +
\frac{43}{600} \langle V \rangle \right. \nonumber\\[1ex]
& & \quad\quad\quad 
  \left. + \frac{(Z\alpha)^4}{n^3} \, F_{\rm IR}(nl_j,Z\alpha) \right]\,,
\end{eqnarray}
where $F_{\rm IR}(nl_j,Z\alpha)$ is a dimensionless function of order
one. The middle-energy part is recovered as
\begin{eqnarray}
\label{DefinitionOfFM}
\Delta E_{\rm M} &=& \frac{\alpha}{\pi} \,
\left[
\frac{279}{200} \langle \beta \rangle +
\frac{219}{200} \langle V \rangle \right. \nonumber\\[1ex]
& & \quad\quad\quad 
  \left. + \frac{(Z\alpha)^4}{n^3} \, F_{\rm M}(nl_j,Z\alpha)
\right]\,,
\end{eqnarray}
and the high-energy part reads~\cite{Mo1974a,Mo1974b}
\begin{eqnarray}
\label{DefinitionOfFH}
\Delta E_{\rm H} &=& \Delta m + \frac{\alpha}{\pi} \,
\left[ - \frac{3}{2} \langle \beta \rangle
- \frac{7}{6} \langle V \rangle \right. \nonumber\\[1ex]
& & \quad\quad\quad\quad\quad\quad
  \left. + \frac{(Z \alpha)^4}{n^3} \, F_{\rm H}(nl_j,Z\alpha)
\right]\,.
\end{eqnarray}
The infrared part is discussed in Sec.~\ref{InfraRed}. The
middle-energy part is divided into a middle-energy subtraction term
$F_{\rm MA}$ and a middle-energy remainder $F_{\rm MB}$. The
subtraction term $F_{\rm MA}$ is discussed in
Sec.~\ref{MiddleEnergySubtraction}, the remainder term $F_{\rm MB}$ is
treated in Sec.~\ref{MiddleEnergyRemainder}. We recover the
middle-energy term as the sum
\begin{equation}
\label{DecompositionOfFM}
F_{\rm M}(nl_j,Z\alpha) \; = \;
F_{\rm MA}(nl_j,Z\alpha) +
F_{\rm MB}(nl_j,Z\alpha)\,.
\end{equation}
A similar separation is employed for the
high-energy part. The high-energy part is divided into a subtraction
term $F_{\rm HA}$, which is evaluated in
Sec.~\ref{HighEnergySubtraction}, and the high-energy remainder
$F_{\rm HB}$, which is discussed in
Sec.~\ref{HighEnergyRemainder}. The sum of the subtraction term and
the remainder is
\begin{equation}
\label{DecompositionOfFH}
F_{\rm H}(nl_j,Z\alpha) \; = \;
F_{\rm HA}(nl_j,Z\alpha) +
F_{\rm HB}(nl_j,Z\alpha)\,.
\end{equation}
The total energy shift is given as
\begin{eqnarray}
\label{net}
\Delta E_{\rm SE} & = &
\Delta E_{\rm IR} + \Delta E_{\rm M} + E_{\rm H} - \Delta m 
\nonumber\\[2ex]
& = & \frac{\alpha}{\pi} \frac{(Z\alpha)^4}{n^3} \,
\left[F_{\rm IR}(nl_j,Z\alpha) \right. \nonumber\\[1ex]
& & \quad\quad\quad 
\left. + F_{\rm M}(nl_j,Z\alpha) + 
F_{\rm H}(nl_j,Z\alpha)\right]\,.
\end{eqnarray}
The scaled self-energy function $F$ defined in Eq.~(\ref{ESEasF}) is 
therefore obtained as
\begin{eqnarray}
F(nl_j,Z\alpha) &=& F_{\rm IR}(nl_j,Z\alpha) \nonumber\\[1ex]
& & \quad\quad 
+ F_{\rm M}(nl_j,Z\alpha) + F_{\rm H}(nl_j,Z\alpha)\,.
\end{eqnarray}
In analogy to the approach described in~\cite{Mo1974a,Mo1982,Mo1992},
we define the low-energy part as the sum of the infrared part and the
middle-energy part,
\begin{eqnarray}
\label{DefinitionOfFL}
\Delta E_{\rm L} & = & \Delta E_{\rm IR} + \Delta E_{\rm M} 
\nonumber\\[1ex]
& = & \frac{\alpha}{\pi}
\left[\frac{3}{2} \langle \beta \rangle +
\frac{7}{6} \langle V \rangle +
\frac{(Z \alpha)^4}{n^3} \, F_{\rm L}(nl_j,Z\alpha)\right]\,,
\end{eqnarray}
where
\begin{equation}
\label{FLasSum}
F_{\rm L}(nl_j,Z\alpha) = 
F_{\rm IR}(nl_j,Z\alpha) +
F_{\rm M}(nl_j,Z\alpha)\,.
\end{equation}
The limits for the functions $F_{\rm L}(nl_j,Z\alpha)$ and 
$F_{\rm H}(nl_j,Z\alpha)$ as $Z\alpha \to 0$ were obtained
in~\cite{Mo1974b,Mo1973,Je1999}.

\typeout{Subsection 2C}
%
%
\subsection{Treatment of the divergent terms}
\label{RegularizationPrescription}

\begin{figure}[htb]
\begin{center}
\begin{minipage}{8.6cm}
\centerline{\mbox{\epsfysize=6.0cm\epsffile{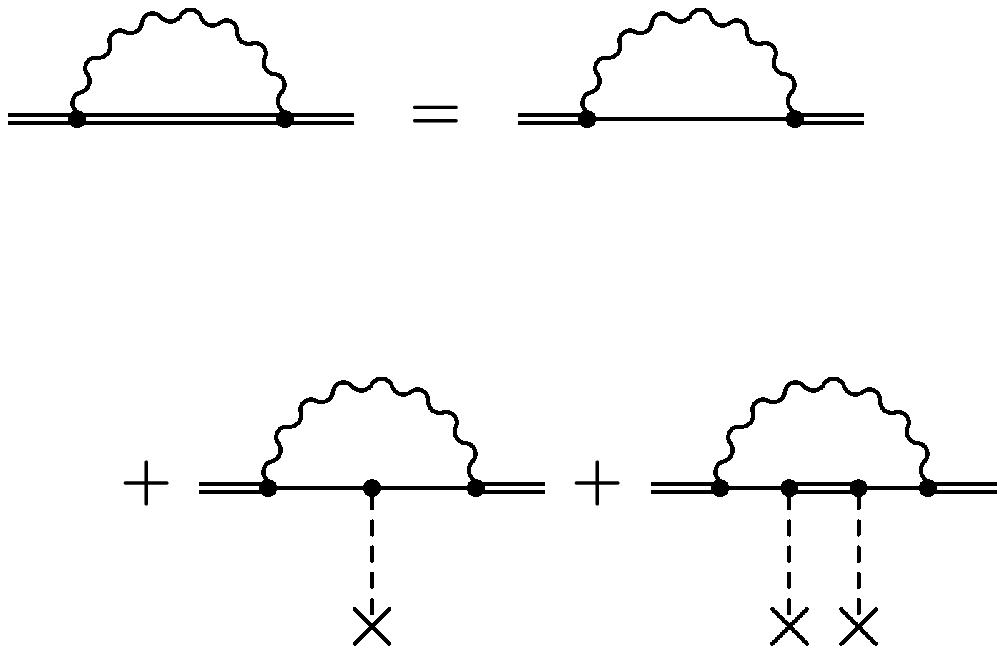}}}
\caption{\label{ExactVExpansion}
The exact expansion of the bound electron propagator in powers of the
binding field leads to a zero-potential, a one-potential, and a
many-potential term.  The dashed lines denote Coulomb photons, the
crosses denote the interaction with the (external) binding field.}
\end{minipage}
\end{center}
\end{figure}

The free electron propagator
\begin{equation}
\label{DefinitionOfF}
F = \frac{1}{\bbox{\alpha} \cdot \bbox{p}
  + \beta - z }
\end{equation}
and the full electron propagator $G$ defined in
Eq.~(\ref{DefinitionOfG}) fulfill the following identity, which is of
particular importance for the validity of the method used in the
numerical evaluation of the all-order binding correction to the Lamb
shift:
\begin{equation}
\label{ExactExpansion}
G = F - F\,V\,F + F\,V\,G\,V\,F\,.
\end{equation}
This identity leads naturally to a separation of the one-photon
self energy into a zero-vertex, a single-vertex, and a many-vertex term.
This is represented diagrammatically in Fig.~\ref{ExactVExpansion}.

%
%

All ultraviolet divergences which occur in the one-photon problem
(mass counter term and vertex divergence) are generated by the
zero-vertex and the single-vertex terms. The many-vertex term is
ultraviolet safe.  Of crucial importance is the observation that one
may additionally simplify the problem by replacing the one-potential
term with an approximate expression in which the potential is
``commuted to the outside.'' The approximate expression generates all
divergences and all terms of lower order than $\alpha\,(Z\alpha)^4$
present in the one-vertex term. Unlike the raw one-potential term, it
is amenable to significant further simplification and can be reduced
to {\em one}-dimensional numerical integrals that can be evaluated
easily (a straightforward formulation of the self-energy problem
requires a {\em three}-dimensional numerical integration).
Without this significant improvement, an all-order calculation would
be much more difficult at low nuclear charge, because the lower-order
terms would introduce significant further numerical cancellations.

In addition, the special approximate resolvent can
be used effectively for an efficient subtraction
scheme in the middle-energy part of the calculation.  In the infrared
part, such a subtraction is not used because it would introduce
infrared divergences.

We now turn to the construction of the special approximate resolvent,
which will be referred to as $G_{\rm A}$ and will be used in this
calculation to isolate the ultraviolet divergences in the high-energy
part (and to provide subtraction terms in the middle-energy
part). It is based on an approximation to the first two terms on the
right-hand side of Eq.~(\ref{ExactExpansion}).  The so-called
one-potential term $FVF$ in Eq.~(\ref{ExactExpansion}) is
approximated by an expression in which the potential terms $V$ are
commuted to the outside:
\begin{equation}
- FVF \approx -\frac{1}{2}\,\left\{V,F^2\right\}\,.
\end{equation}
Furthermore, the following identity is used:
\begin{eqnarray}
F^2 &=& 
  \left( \frac{1}{\bbox{\alpha}\cdot\bbox{p} + \beta - z} \right)^2 
\nonumber\\[1ex]
&=& \frac{1}{\bbox{p}^2 + 1 - z^2} +
\frac{2\,z\,(\beta + z)}{\left(\bbox{p}^2 + 1 - z^2\right)^2} 
\nonumber\\[1ex]
& & \quad\quad\quad\quad\quad\quad\quad\quad\quad
  + \frac{2\,z\,(\bbox{\alpha}\cdot\bbox{p})}
  {\left(\bbox{p}^2 + 1 - z^2\right)^2}\,.
\end{eqnarray}
In $2\times2$ spinor space, this expression may be divided into a
diagonal and a non-diagonal part. The diagonal part is
\begin{equation}
{\rm diag}(F^2) = \frac{1}{\bbox{p}^2 + 1 - z^2} +
\frac{2\,z\,(\beta + z)}{\left(\bbox{p}^2 + 1 - z^2\right)^2}\,.
\end{equation}
The off-diagonal part is given by
\[
F^2 - {\rm diag}(F^2) = \frac{2\,z\,(\bbox{\alpha}\cdot\bbox{p})}
  {\left(\bbox{p}^2 + 1 - z^2\right)^2}\,.
\]
We define the resolvent $G_{\rm A}$ as
\begin{equation}
\label{DefinitionOfGA}
G_{\rm A} = F - \frac{1}{2}\,\left\{V,{\rm diag}\left(F^2\right)\right\}\,.
\end{equation}
All divergences which occur in the self energy are generated by the
simplified propagator $G_{\rm A}$.  We define the propagator $G_{\rm
B}$ as the difference of $G$ and $G_{\rm A}$,
\begin{eqnarray}
\label{DefinitionOfGB}
G_{\rm B} &=& G - G_{\rm A} \nonumber\\[1ex]
&=& \frac{1}{2}\,\left\{V,{\rm diag}(F^2)\right\} - F\,V\,F 
    + F\,V\,G\,\,V\,F\,.
\end{eqnarray}
$G_{\rm B}$ does not generate any divergences and leads to the
middle-energy remainder discussed in Sec.~\ref{MiddleEnergyRemainder}
and the high-energy remainder (Sec.~\ref{HighEnergyRemainder}).

\typeout{Section 3}
%
%
\section{The Low-Energy Part}
\label{LowEnergyPart}

\typeout{Subsection 3A}
%
%
\subsection{The Infrared Part}
\label{InfraRed}

The infrared part is given by
\begin{eqnarray}
\label{deltaEIR}
\lefteqn{\Delta E_{\rm IR} = 
{\rm -i}\, e^2 \, {\rm Re} \,
\int_{C_{\rm IR}} \! \frac{{\rm d}\omega}{2 \pi}
\int \! \! \frac{{\rm d}^3 \bbox{k}}{(2 \pi)^3} \,
D_{\mu\nu}(k^2)} \nonumber\\[2ex]
& & \times \left< \psi \left| \, \alpha^{\mu} \,
{\rm e}^{{\rm i}\,\bbox{k}\cdot\bbox{x}}\,
G(E_n - \omega) \, \alpha^{\nu} \, 
{\rm e}^{-{\rm i}\,\bbox{k}\cdot\bbox{x}}\,
\right| \psi \right> \,,
\end{eqnarray}
where relevant definitions of the symbols can be found in
Eqs.~(\ref{deltaESEM}--\ref{PhotonPropagator}), the contour $C_{\rm
IR}$ is as shown in Fig.~\ref{SeparationLow}, and the unregularized
version of the photon propagator
\begin{equation}
\label{PhotonPropagatorUnreg}
D_{\mu\nu}(k^2) = 
- \frac{\displaystyle g_{\mu\nu}}{\displaystyle k^2 + {\rm i}\,\epsilon} 
\end{equation}
may be used. The infrared part consists of the following integration
region for the virtual photon:
\begin{equation}
\left.
\begin{array}{l}
\omega \in \left(0, \lfrac{1}{10}\,E_n\right) \pm i\,\delta \\[1ex]  
z \in \left(\lfrac{9}{10}\,E_n, E_n\right) \pm i\,\delta
\end{array}
\right\}
\mbox{infrared part $\Delta E_{\rm IR}$}\,.
\end{equation}
Following Secs.~2 and 3 of~\cite{Mo1974a}, we write $\Delta E_{\rm
IR}$ as a three-dimensional integral [see, e.g., Eqs.~(3.4), (3.11),
and (3.14) of~\cite{Mo1974a}]
\begin{eqnarray}
\label{defEIR}
\Delta E_{\rm IR} &=& \frac{\alpha}{\pi} \, \frac{E_n}{10} 
- \frac{\alpha}{\pi} \, ({\rm P. V.}) \, 
\int_{\lfrac{9}{10}\,E_n}^{E_n} {\rm d}z 
\nonumber\\[1ex]
& &
\int_0^\infty {\rm d} x_1 \, x_1^2 
\int_0^\infty {\rm d} x_2 \, x_2^2 \,\, {\cal M}_{\rm IR}(x_2,x_1,z)\,,
\end{eqnarray}
where
\begin{eqnarray}
\label{defMIR}
& & {\cal M}_{\rm IR}(x_2,x_1,z) =
\sum_\kappa \sum_{i,j=1}^{2} 
\nonumber\\[2ex]
& & \quad\quad f_{\bar \imath}(x_2) \,
G^{ij}_{\kappa}(x_2,x_1,z) \, 
f_{\bar \jmath}(x_1) \, 
A^{ij}_{\kappa}(x_2,x_1)\,.
\end{eqnarray}
Here, the quantum number $\kappa$ is the Dirac angular quantum number
of the intermediate state, 
\begin{equation}
\label{DefOfkappa}
\kappa = 2 \, (l-j) \, (j+1/2)\,,
\end{equation}
where $l$ is the orbital angular momentum quantum number and $j$ is
the total angular momentum of the bound electron. The functions
$f_i(x_2)$ ($i=1,2$) are the radial wave functions defined in
Eq.~(A.4) in~\cite{Mo1974a} for an arbitrary bound state (and in
Eq.~(A.8) in~\cite{Mo1974a} for the 1S state). We define 
${\bar \imath} = 3 - i$. The functions $G^{ij}_{\kappa}(x_2,x_1,z)$ ($i,j=1,2$)
are the radial Green functions, which result from a decomposition of
the electron Green function defined in Eq.~(\ref{DefinitionOfG}) into
partial waves.  The explicit formulas are given in Eq.~(A.16)
in~\cite{Mo1974a}.

The photon angular functions $A^{ij}_{\kappa}$ ($i,j=1,2$) are defined
in Eq.~(3.15) of Ref.~\cite{Mo1974a} for an arbitrary bound state. In
Eq.~(3.17) in~\cite{Mo1974a}, specific formulas are given for the 1S
state.  In Eqs.~(2.2), (2.3) and (2.4) of~\cite{Mo1982}, the
special cases of ${\rm S}_{1/2}$, ${\rm P}_{1/2}$ and ${\rm P}_{3/2}$
states are considered. Further relevant formulas for excited states
can be found in~\cite{MoKi1992}. The photon angular functions depend
on the energy argument $z$, but this dependence is usually
suppressed. The summation over $\kappa$ in Eq.~(\ref{defMIR}) extends
over all negative and all positive integers, excluding zero.  We
observe that the integral is symmetric under the interchange of the
radial coordinates $x_2$ and $x_1$, so that
\begin{eqnarray}
\label{IntegralEIR}
\Delta E_{\rm IR} &=& \frac{\alpha}{\pi} \, \frac{E_n}{10} - 
\frac{2\,\alpha}{\pi} \, ({\rm P. V.}) 
\int_{\lfrac{9}{10}\,E_n}^{E_n} {\rm d}z 
\nonumber\\[2ex]
& &
\int_0^\infty {\rm d} x_1 \, x_1^2 
\int_0^{x_1} {\rm d} x_2 \, x_2^2 \,\, {\cal M}_{\rm IR}(x_2,x_1,z)\,.
\end{eqnarray}
The following variable substitution,
\begin{equation}
\label{DefinitionOfR}
r = x_2/x_1\,,\;\;\; y = a\,x_1\,,
\end{equation}
is made, so that $r \in (0,1)$ and $y \in (0,\infty)$.  The scaling
variable $a$ is defined as
\begin{equation}
\label{DefinitionOfA}
a = 2 \,\sqrt{1-E_n^2}\,.
\end{equation}
The Jacobian is
\begin{equation}
\left|\frac{\partial(x_2,x_1)}{\partial(r,y)}\right| =
\left| \begin{array}{cc}
\frac{\textstyle \partial x_2}{\textstyle \partial r} & 
\frac{\textstyle \partial x_1}{\textstyle \partial r} \\[2ex]
\frac{\textstyle \partial x_2}{\textstyle \partial y} & 
\frac{\textstyle \partial x_1}{\textstyle \partial y} 
\end{array} \right| = \frac{y}{a^2}\,.
\end{equation}
The function $S_{\rm IR}$ is given by
\begin{eqnarray}
\label{defSIR}
S_{\rm IR}(r,y,z) &=& - \frac{2 \, r^2 \, y^5}{a^6} \, 
{\cal M}_{\rm IR}\left(\frac{r\,y}{a},\frac{y}{a},z\right) \nonumber\\[2ex]
&=& - \frac{2 \, r^2 \, y^5}{a^6} \, 
\sum_{|\kappa|=1}^{\infty} \sum_{\kappa = \pm |\kappa|} 
\sum_{i,j=1}^{2} f_{\bar \imath}\left(\frac{r\,y}{a}\right) \nonumber\\[1ex]
& & \, \times \,\,
G^{ij}_{\kappa}\left(\frac{r\,y}{a},\frac{y}{a},z\right) \, 
f_{\bar \jmath}\left(\frac{y}{a}\right) \, 
A^{ij}_{\kappa}\left(\frac{r\,y}{a},\frac{y}{a}\right) \nonumber\\[2ex]
&=& - \frac{2 \, r^2 \, y^5}{a^6} \,  
\sum_{|\kappa|=1}^{\infty} T_{{\rm IR},|\kappa|}(r,y,z)\,,
\end{eqnarray}
where in the last line we define implicitly the terms $T_{{\rm
IR},|\kappa|}$ for $|\kappa| = 1,\dots,\infty$ as
\begin{eqnarray}
\label{DefinitionOfTIRkappa}
& & T_{{\rm IR},|\kappa|}(r,y,z) =
\sum_{\kappa = \pm |\kappa|} 
\sum_{i,j=1}^{2} \nonumber\\[1ex]
& & \quad f_{\bar \imath}\left(\frac{r\,y}{a}\right) \,
G^{ij}_{\kappa}\left(\frac{r\,y}{a},\frac{y}{a},z\right) \, 
f_{\bar \jmath}\left(\frac{y}{a}\right) \, 
A^{ij}_{\kappa}\left(\frac{r\,y}{a},\frac{y}{a}\right)\,.
\end{eqnarray}
Using the definition~(\ref{defSIR}), we obtain for $\Delta E_{\rm IR}$,
\begin{eqnarray}
\label{EvaluationOfEIR}
\Delta E_{\rm IR} &=& \frac{\alpha}{\pi}\,\frac{E_n}{10} + 
\frac{\alpha}{\pi} \, ({\rm P.V.})
\int_{\lfrac{9}{10}\,E_n}^{E_n} {\rm d}z 
\nonumber\\[1ex]
& & \int_0^1 {\rm d}r \, \int_0^\infty {\rm d}y \, 
S_{\rm IR}(r,y,z)\,.
\end{eqnarray}
The specification of the principal value (P.V.) is necessary for the
excited states of the L shell, because of the poles along the
integration contour which correspond to the spontaneous decay into the
ground state.  Here we are exclusively concerned with the real part of
the energy shift, as specified in
Eq.~(\ref{deltaEIR}), which is equivalent to the specification of the
principal value in~(\ref{EvaluationOfEIR}). 
Evaluation of the integral over $z$ is facilitated by the
subtraction of those terms which generate the singularities along the
integration contour (for higher excited states, there can be numerous
bound state poles, as pointed out in~\cite{MoKi1992,JeSoMo1997}). For
the 2S and $2{\rm P}_{1/2}$ states, only the pole contribution from
the ground state must be subtracted. For the $2{\rm P}_{3/2}$ state,
pole contributions originating from the 1S, the 2S and the $2{\rm
P}_{1/2}$ states must be taken into account. The numerical evaluation
of the subtracted integrand proceeds along ideas outlined
in~\cite{Mo1982,MoKi1992} and is not discussed here in any further
detail.

The scaling parameter $a$ for the integration over $y$ is chosen to
simplify the exponential dependence of the function $S$ defined in
Eq.~(\ref{defSIR}). The main exponential dependence is given by the
relativistic radial wave functions (upper and lower components).  Both
components [$f_1(x)$ and $f_2(x)$] vary approximately as (neglecting
relatively slowly varying factors)
\[
\exp\left(-a\,x/2\right) 
\;\;\;\;\;\;\;\; \mbox{(for large $x$)}\,.
\]
The scaling variable $a$, expanded in powers of $Z\alpha$, is
\begin{eqnarray}
\label{ExpansionOfA}
a &=& 2\,\sqrt{1-E_n^2} \nonumber\\[2ex]
&=& 2\,\sqrt{1 - \left(1 - \frac{(Z\alpha)^2}{2\,n^2} + 
{\rm O}\left[(Z\alpha)^4\right]\right)^2} \nonumber\\[2ex]
&=& 2\,\frac{Z\alpha}{n} + {\rm O}\!\left[(Z\alpha)^3\right]\,.
\end{eqnarray}
Therefore, $a$ is just twice the {\em inverse} of the Bohr radius
$n/(Z\alpha)$ in the nonrelativistic limit. The product
\[
f_{\bar \imath}\left(\frac{r y}{a}\right) \times
f_{\bar \jmath}\left(\frac{y}{a}\right)
\;\;\;\;\;\;\;\;\; \mbox{for arbitrary ${\bar \imath},{\bar \jmath} \in \{1,2\}$}
\]
[which occurs in Eq.~(\ref{defSIR})] depends on the radial arguments
approximately as
\[
e^{-y} \times \exp\left[\lfrac{1}{2}\,(1-r)\,y\right] 
\;\;\;\;\;\;\;\;\; \mbox{(for large $y$)}\,.
\]
Note that the main dependence as given by the term $\exp(-y)$ is
exactly the weight factor of the Gau\ss{}-Laguerre integration
quadrature formula. The deviation from the exact $\exp(-y)$--type
behavior becomes smaller as $r \to 1$. This is favorable because the
region near $r = 1$ gives a large contribution to the integral
in~(\ref{EvaluationOfEIR}).

\widetext

%
%
\begin{table}[thb]
\begin{center}
\begin{minipage}{17cm}
\begin{center}
\caption{\label{tableFIRKL} Infrared part for the K and
L shell states,
$F_{\rm IR}(1{\rm S}_{1/2},Z\alpha)$,
$F_{\rm IR}(2{\rm S}_{1/2},Z\alpha)$,
$F_{\rm IR}(2{\rm P}_{1/2},Z\alpha)$,
and $F_{\rm IR}(2{\rm P}_{3/2},Z\alpha)$,
evaluated for low-$Z$
hydrogenlike ions. The calculations were performed with the 
numerical value of $\alpha^{-1} = 137.036$
for the fine-structure constant.}
\begin{tabular}{lr@{.}lr@{.}lr@{.}lr@{.}l}
\multicolumn{1}{c}{$Z$} &
\multicolumn{2}{c}
  {\rule[-3mm]{0mm}{8mm}$F_{\rm IR}(1{\rm S}_{1/2},Z\alpha)$} &
\multicolumn{2}{c}
  {\rule[-3mm]{0mm}{8mm}$F_{\rm IR}(2{\rm S}_{1/2},Z\alpha)$} &
\multicolumn{2}{c}
  {\rule[-3mm]{0mm}{8mm}$F_{\rm IR}(2{\rm P}_{1/2},Z\alpha)$} &
\multicolumn{2}{c}
  {\rule[-3mm]{0mm}{8mm}$F_{\rm IR}(2{\rm P}_{3/2},Z\alpha)$}
\\
\hline
$1$ & 
 $7$ & $236~623~736~8(1)$ & 
 $7$ & $479~764~180(1)$ &
 $0$ & $085~327~852(1)$ &
 $0$ & $082~736~497(1)$ \\
$2$  &
 $5$ & $539~002~119~1(1)$ & 
 $5$ & $782~025~637(1)$ &
 $0$ & $086~073~669(1)$ &
 $0$ & $083~279~461(1)$ \\
$3$  &
 $4$ & $598~155~821~8(1)$ &
 $4$ & $840~923~962(1)$ &
 $0$ & $087~162~510(1)$ &
 $0$ & $084~091~830(1)$ \\
$4$  &
 $3$ & $963~124~140~6(1)$ &
 $4$ & $205~501~798(1)$ &
 $0$ & $088~543~188(1)$ &
 $0$ & $085~140~788(1)$ \\
$5$  &
 $3$ & $493~253~319~4(1)$ & 
 $3$ & $735~114~958(1)$ &
 $0$ & $090~180~835(1)$ &
 $0$ & $086~403~178(1)$ \\
\end{tabular}
\end{center}
\end{minipage}
\end{center}
\end{table}

\narrowtext

The sum over $|\kappa|$ in Eq.~(\ref{defSIR}) is carried out locally,
i.e.,~for each set of arguments $r,y,z$. The sum over $|\kappa|$ is
absolutely convergent. For $|\kappa| \to \infty$, the convergence of
the sum is governed by the asymptotic behavior of the Bessel functions
which occur in the photon functions $A^{ij}_{\kappa}$ ($i,j=1,2$) [see
Eqs.~(3.15) and (3.16) in~\cite{Mo1974a}]. The photon functions
contain products of two Bessel functions of the form ${\cal
J}_l(\rho_{2/1})$ where ${\cal J}_l$ stands for either $j_l$ or
$j'_l$, and the index $l$ is in the range
$l~\in~\{|\kappa|-1,|\kappa|,|\kappa| + 1\}$. The argument is either
$\rho_2 = (E_n - z)\,x_2$ or $\rho_1 = (E_n - z)\,x_1$. The asymptotic
behavior of the two relevant Bessel functions for large $l$ (and
therefore large $|\kappa|$) is
\begin{eqnarray}
\label{BesselAsymptotics1}
j'_l(x) & = & 
\frac{l}{x}\,\frac{x^l}{(2 l + 1)!!} \,
\left[1 + {\rm O}\left( \frac{1}{l} \right) \right]
\quad \mbox{and}\\
\label{BesselAsymptotics2}
j_l(x) & = & 
\frac{x^l}{(2 l + 1)!!} \,
\left[1 + {\rm O}\left( \frac{1}{l} \right) \right]\,.
\end{eqnarray}
This implies that when $\min\{\rho_2,\rho_1\} = \rho_2 < l$, the
function ${\cal J}_l(\rho_2)$ vanishes with increasing $l$
approximately as $({\rm e}\,\rho_2/2 l)^l$. This rapidly converging asymptotic
behavior sets in as soon as $l \approx |\kappa| > \rho_2 =
r\,\omega\,y/a$ [see Eqs.~(\ref{DefOfZ}) 
and~(\ref{DefinitionOfTIRkappa})].
Due to the rapid convergence for $|\kappa| > \rho_2$, the maximum
angular momentum quantum number $|\kappa|$ in the numerical
calculation of the infrared part is less than $3~000$. Note that
because $z~\in~{\bf (}\lfrac{9}{10}\,E_n,E_n{\bf )}$ in the infrared
part, $\omega < \lfrac{1}{10}\,E_n$.

The integration scheme is based on a crude estimate of the dependence
of the integrand $S_{\rm IR}(r,y,z)$ defined in~Eq.~(\ref{defSIR}) on
the integration variables $r$, $y$ and $z$. The main contribution to
the integral is given by the region where the arguments of the
Whittaker functions as they occur in the Green function~[see
Eq.~(A.16)~in~\cite{Mo1974a}] are much larger than the Dirac angular
momentum,
\[
2\,c\,\frac{y}{a} \gg |\kappa|
\]
(see also p.~56 of~\cite{Mo1974b}).  We assume the asymptotic form of
the Green function given in Eq.~(A.3) in~\cite{Mo1974b} 
applies and attribute a factor
\[
\exp[-(1-r)\,c\,y/a]
\]
to the radial Green functions $G^{ij}_{\kappa}$ as they occur in
Eq.~(\ref{defSIR}). Note that relatively slowly varying factors are
replaced by unity. The products of the radial wave functions $f_{\bar
\imath}$ and $f_{\bar \jmath}$, according to the discussion following
Eq.~(\ref{ExpansionOfA}), behave as
\[
e^{-y} \, \exp\left[\lfrac{1}{2}\,(1-r)\,y\right]
\]
for large $y$. The photon functions $A^{ij}_{\kappa}$ in
Eq.~(\ref{defSIR}) give rise to an approximate factor
\begin{equation}
\frac{\sin[(1-r)\,(E_n-z)\,y/a]}{(1-r)}\,.
\end{equation}
Therefore [see also Eq.~(2.12) in~\cite{Mo1974b}], we base our
choice of the integration routine on the approximation
\begin{eqnarray}
\label{approxsir}
e^{-y} \, \exp\left[-\left(\frac{c}{a} - 
  \frac{1}{2}\right)\,(1-r)\,y\right] 
\nonumber\\[2ex]
\quad\quad
  \times \frac{\sin\left[(1-r)\,(E_n-z)\,y/a\right]}{(1-r)}
\end{eqnarray}
for $S_{\rm IR}$. The three-dimensional integral in~(\ref{EvaluationOfEIR}) 
is evaluated by successive Gaussian
quadrature.  Details of the integration procedure can be found
in~\cite{Je1999}.

In order to check the numerical stability of the results, the
calculations are repeated with three different values of the 
fine-structure constant $\alpha$:
\begin{equation}
\label{DefinitionOfAlphas}
\begin{array}{rcll}
\alpha_< & = & 1/137.036~000~5 \,, & \\
\alpha_0 & = & 1/137.036~000~0 & \mbox{and,}\\
\alpha_> & = & 1/137.035~999~5 \,. &
\end{array}
\end{equation}
These values are close to the 1998 CODATA recommended value of
$\alpha^{-1} = 137.035~999~76(50)$~\cite{MoTa2000}. The calculation
was parallelized using the Message Passing Interface (MPI) and carried
out on a cluster of Silicon Graphics workstations and on an IBM 9276
SP/2 multiprocessor system~\cite{Disclaimer}. The results for the
infrared part $F_{\rm IR}$, defined in Eq.~(\ref{DefinitionOfFIR}), are
given in~Table~\ref{tableFIRKL} for a value of $\alpha^{-1} =
\alpha_0^{-1} = 137.036$. This value of $\alpha$ will be used
exclusively in the numerical evaluations presented here. For numerical
results obtained by employing the values of $\alpha_<$ and $\alpha_>$
[see Eq.~(\ref{DefinitionOfAlphas})] we refer to~\cite{Je1999}.

\typeout{Subsection 3B}
%
%
\subsection{The Middle-Energy Subtraction Term}
\label{MiddleEnergySubtraction}

The middle-energy part is given by
\begin{eqnarray}
\label{deltaEM}
\lefteqn{\Delta E_{\rm M} = 
{\rm -i}\, e^2 \, 
\int_{C_{\rm M}} \! \frac{{\rm d}\omega}{2 \pi}
\int \! \! \frac{{\rm d}^3 \bbox{k}}{(2 \pi)^3} \,
D_{\mu\nu}(k^2)} \nonumber\\[2ex]
& & \times \left< \psi \left| \, \alpha^{\mu} \,
{\rm e}^{{\rm i}\,\bbox{k}\cdot\bbox{x}}\,
G(E_n - \omega) \, \alpha^{\nu} \, 
{\rm e}^{-{\rm i}\,\bbox{k}\cdot\bbox{x}}\,
\right| \psi \right> \,,
\end{eqnarray}
where relevant definitions of the symbols can be found in
Eqs.~(\ref{deltaESEM})--(\ref{PhotonPropagator})
and Eq.~(\ref{PhotonPropagatorUnreg}), and
the contour $C_{\rm M}$ is as shown in Fig.~\ref{SeparationLow}.
The middle-energy part consists of 
the following integration region for the
virtual photon:
\begin{equation}
\left.
\begin{array}{l}
\omega \in \left(\lfrac{1}{10}\,E_n,E_n\right) \pm {\rm i}\,\delta \\[1ex]  
z \in \left(0,\lfrac{9}{10}\,E_n\right) \pm {\rm i}\,\delta
\end{array}
\right\}
\mbox{middle-energy part $\Delta E_{\rm M}$}\,.
\end{equation}
The numerical evaluation of the middle-energy part is simplified
considerably by the decomposition of the relativistic Dirac-Coulomb
Green function $G$ as
\begin{equation}
\label{DecompositionOfG}
G \;\; = \;\; G_{\rm A} \; + \; G_{\rm B}\,,
\end{equation}
where $G_{\rm A}$ is defined in (\ref{DefinitionOfGA}) and represents
the sum of an approximation to the so-called zero- and one-potential
terms generated by the expansion of the Dirac-Coulomb Green function
$G$ in powers of the binding field $V$.  We define the middle-energy
subtraction term $F_{\rm MA}$ as the expression obtained upon
substitution of the propagator $G_{\rm A}$ for $G$ in
Eq.~(\ref{deltaEM}).  The propagator $G_{\rm B}$ is simply calculated
as the difference of $G$ and $G_{\rm A}$ [see
Eq.~(\ref{DefinitionOfGB})]. A substitution of the propagator $G_{\rm
B}$ for $G$ in Eq.~(\ref{deltaEM}) leads to the middle-energy
remainder $F_{\rm MB}$ which is discussed in
Sec.~\ref{MiddleEnergyRemainder}. We provide here the explicit
expressions
\begin{eqnarray}
\label{deltaEMA}
\lefteqn{\Delta E_{\rm MA} =
{\rm -i}\, e^2 \,
\int_{C_{\rm M}} \! \frac{{\rm d}\omega}{2 \pi}
\int \! \! \frac{{\rm d}^3 \bbox{k}}{(2 \pi)^3} \,
D_{\mu\nu}(k^2)} \nonumber\\[2ex]
& & \times \left< \psi \left| \, \alpha^{\mu} \,
{\rm e}^{{\rm i}\,\bbox{k}\cdot\bbox{x}}\,
G_{\rm A}(E_n - \omega) \, \alpha^{\nu} \, 
{\rm e}^{-{\rm i}\,\bbox{k}\cdot\bbox{x}}\, \right| \psi \right> 
\end{eqnarray}
and
\begin{eqnarray}
\label{deltaEMB}
\lefteqn{\Delta E_{\rm MB} =
{\rm -i}\, e^2 \,
\int_{C_{\rm M}} \! \frac{{\rm d}\omega}{2 \pi}
\int \! \! \frac{{\rm d}^3 \bbox{k}}{(2 \pi)^3} \,
D_{\mu\nu}(k^2)} \nonumber\\[2ex]
& & \times \left< \psi \left| \, \alpha^{\mu} \,
{\rm e}^{{\rm i}\,\bbox{k}\cdot\bbox{x}}\,
G_{\rm B}(E_n - \omega) \, \alpha^{\nu} \,
{\rm e}^{-{\rm i}\,\bbox{k}\cdot\bbox{x}}\,
\right| \psi \right> \,.
\end{eqnarray}
Note that the decomposition of the Dirac-Coulomb Green function as
in~(\ref{DecompositionOfG}) is not applicable in the infrared part,
because of numerical problems for ultra-soft photons (infrared
divergences). Rewriting (\ref{deltaEMA}) appropriately
into a three-dimensional integral~\cite{Mo1974a,Mo1974b,Je1999},
we have
\begin{eqnarray}
\label{IntegralEMA}
\lefteqn{\Delta E_{\rm MA} = \frac{\alpha}{\pi} \, \frac{9}{10} \, E_n - 
\frac{2\,\alpha}{\pi} \, 
\int_{0}^{\lfrac{9}{10}\,E_n} \!\!\!\! {\rm d}z }
\nonumber\\[2ex] 
& & \int_0^\infty \! {\rm d} x_1 \, x_1^2 \,
\int_0^{x_1} \! {\rm d} x_2 \, x_2^2 \,\, {\cal M}_{\rm MA}(x_2,x_1,z)\,.
\end{eqnarray}
The function ${\cal M}_{\rm MA}(x_2,x_1,z)$ is defined in analogy to
the function ${\cal M}_{\rm IR}(x_2,x_1,z)$ defined in
Eq.~(\ref{defMIR}) for the infrared part. Also, we define a function
${\cal S}_{\rm MA}(x_2,x_1,z)$ in analogy to the function ${\cal
S}_{\rm IR}(x_2,x_1,z)$ given in Eq.~(\ref{defSIR}) for the infrared
part, which will be used in Eq.~(\ref{DefinitionOfEMA}) below.
We have,
\begin{eqnarray}
\label{defSMA}
S_{\rm MA}(r,y,z) &=& - \frac{2 \, r^2 \, y^5}{a^6} \, 
{\cal M}_{\rm MA}\left(\frac{r\,y}{a},\frac{y}{a},z\right) \nonumber\\[2ex]
&=& - \frac{2 \, r^2 \, y^5}{a^6} \, 
\sum_{|\kappa|=1}^{\infty} \sum_{\kappa = \pm |\kappa|} 
\sum_{i,j=1}^{2} f_{\bar \imath}\left(\frac{r\,y}{a}\right) 
\nonumber\\[1ex]
& & \, \times \,\,
G^{ij}_{{\rm A},\kappa}\left(\frac{r\,y}{a},\frac{y}{a},z\right) \, 
f_{\bar \jmath}\left(\frac{y}{a}\right) \, 
A^{ij}_{\kappa}\left(\frac{r\,y}{a},\frac{y}{a}\right) \nonumber\\[2ex]
&=& - \frac{2 \, r^2 \, y^5}{a^6} \,
\sum_{|\kappa|=1}^{\infty} T_{{\rm MA},|\kappa|}(r,y,z)\,.
\end{eqnarray}
The expansion of the propagator $G_{\rm A}$ into partial waves is
given in Eqs.~(5.4) and (A.20) in \cite{Mo1974a} and in Eqs.~(D.37) and (D.42)
in~\cite{Je1999}. This expansion leads to the component
functions $G^{ij}_{{\rm A},\kappa}$. The terms $T_{{\rm MA},|\kappa|}$ in
the last line of Eq.~(\ref{defSMA}) read
\begin{eqnarray}
\label{DefinitionOfTMAkappa}
& & T_{{\rm MA},|\kappa|}(r,y,z) = \sum_{\kappa = \pm |\kappa|} 
\sum_{i,j=1}^{2} \nonumber\\[1ex]
& & f_{\bar \imath}\left(\frac{r\,y}{a}\right) \,
G^{ij}_{{\rm A},\kappa}\left(\frac{r\,y}{a},\frac{y}{a},z\right) \, 
f_{\bar \jmath}\left(\frac{y}{a}\right) \, 
A^{ij}_{\kappa}\left(\frac{r\,y}{a},\frac{y}{a}\right)\,.
\end{eqnarray}
With these definitions, the middle-energy subtraction term $\Delta
E_{\rm MA}$ can be written as
\begin{eqnarray}
\label{DefinitionOfEMA}
\Delta E_{\rm MA} &=& \frac{\alpha}{\pi}\,\frac{9}{10}\,E_n 
+ \frac{\alpha}{\pi} \, \int_0^{\lfrac{9}{10}\,E_n} \!\!\!\! {\rm d}z
\nonumber\\[2ex]
& & \quad 
\int_0^\infty \! {\rm d}y \, 
\int_0^1 \! {\rm d}r \,\, 
S_{\rm MA}(r,y,z)\,.
\end{eqnarray}
The subtracted lower-order terms yield,
\begin{eqnarray}
\label{DefinitionOfFMA}
\Delta E_{\rm MA} &=& \frac{\alpha}{\pi} \,
\left[
\frac{279}{200} \langle \beta \rangle +
\frac{219}{200} \langle V \rangle \right.
\nonumber\\[1ex]
& & \quad\quad\quad\quad\quad\quad
 \left. + \frac{(Z\alpha)^4}{n^3}\,F_{\rm MA}(nl_j,Z\alpha)
\right]\,.
\end{eqnarray}
The three-dimensional integral in (\ref{DefinitionOfEMA}) is evaluated 
by successive Gaussian quadrature. Details of the 
integration procedure can be found in~\cite{Je1999}.
The numerical results are summarized in
Table~\ref{tableFMKL}.

\widetext

%
%
\begin{table}[thb]
\begin{center}
\begin{minipage}{14cm}
\begin{center}
\caption{\label{tableFMKL} Numerical results for the
middle-energy subtraction term $F_{\rm MA}$,
the middle-energy remainder term $F_{\rm MB}$,
and the middle-energy term $F_{\rm M}$.
The middle-energy term $F_{\rm M}$ is given as the sum 
$F_{\rm M}(nl_j,Z\alpha) = 
F_{\rm MA}(nl_j,Z\alpha) + F_{\rm MB}(nl_j,Z\alpha)$
[see also Eqs.(\ref{DefinitionOfFM}),
(\ref{DefinitionOfFMA}), and (\ref{DefinitionOfFMB})].}
\begin{tabular}{lr@{.}lr@{.}lr@{.}lr@{.}l}
\multicolumn{1}{c}{$Z$} &
\multicolumn{2}{c}
  {\rule[-3mm]{0mm}{8mm}$F_{\rm MA}(1{\rm S}_{1/2},Z\alpha)$} &
\multicolumn{2}{c}
  {\rule[-3mm]{0mm}{8mm}$F_{\rm MA}(2{\rm S}_{1/2},Z\alpha)$} &
\multicolumn{2}{c}
  {\rule[-3mm]{0mm}{8mm}$F_{\rm MA}(2{\rm P}_{1/2},Z\alpha)$} &
\multicolumn{2}{c}
  {\rule[-3mm]{0mm}{8mm}$F_{\rm MA}(2{\rm P}_{3/2},Z\alpha)$}
\\
\hline
$1$  &
 $2$ & $699~379~904~5(1)$ &
 $2$ & $720~878~318(1)$ &
 $0$ & $083~207~314(1)$ &
 $0$ & $701~705~240(1)$ \\
$2$  &
 $2$ & $659~561~381~1(1)$ &
 $2$ & $681~820~660(1)$ &
 $0$ & $084~208~832(1)$ &
 $0$ & $701~850~024(1)$ \\
$3$  &
 $2$ & $623~779~453~0(1)$ &
 $2$ & $647~262~568(1)$ &
 $0$ & $085~831~658(1)$ &
 $0$ & $702~091~147(1)$ \\
$4$  &
 $2$ & $591~151~010~1(1)$ &
 $2$ & $616~290~432(1)$ &
 $0$ & $088~040~763(1)$ &
 $0$ & $702~426~850(1)$ \\
$5$  & 
 $2$ & $561~096~522~1(1)$ &
 $2$ & $588~297~638(1)$ &
 $0$ & $090~803~408(1)$ &
 $0$ & $702~854~461(1)$ \\
\hline
\hline
\multicolumn{1}{c}{$Z$} &
\multicolumn{2}{c}
  {\rule[-3mm]{0mm}{8mm}$F_{\rm MB}(1{\rm S}_{1/2},Z\alpha)$} &
\multicolumn{2}{c}
  {\rule[-3mm]{0mm}{8mm}$F_{\rm MB}(2{\rm S}_{1/2},Z\alpha)$} &
\multicolumn{2}{c}
  {\rule[-3mm]{0mm}{8mm}$F_{\rm MB}(2{\rm P}_{1/2},Z\alpha)$} &
\multicolumn{2}{c}
  {\rule[-3mm]{0mm}{8mm}$F_{\rm MB}(2{\rm P}_{3/2},Z\alpha)$}
\\
\hline
$1$  &
 $ 1$ & $685~993~923~2(1)$ & 
 $ 1$ & $784~756~705(2)$ &
 $ 0$ & $771~787~771(2)$ &
 $-0$ & $094~272~681(2)$ \\
$2$  &
 $ 1$ & $626~842~294~5(1)$ & 
 $ 1$ & $725~583~798(2)$ &
 $ 0$ & $770~778~394(2)$ &
 $-0$ & $094~612~071(2)$ \\
$3$  &
 $ 1$ & $571~406~090~7(1)$ & 
 $ 1$ & $670~086~996(2)$ &
 $ 0$ & $769~153~314(2)$ &
 $-0$ & $095~165~248(2)$ \\
$4$  &
 $ 1$ & $519~082~768~6(1)$ & 
 $ 1$ & $617~650~004(2)$ &
 $ 0$ & $766~954~435(2)$ &
 $-0$ & $095~922~506(2)$ \\
$5$  &
 $ 1$ & $469~482~409~0(1)$ & 
 $ 1$ & $567~873~140(2)$ &
 $ 0$ & $764~220~149(2)$ &
 $-0$ & $096~874~556(2)$ \\
\hline
\hline
\multicolumn{1}{c}{$Z$} &
\multicolumn{2}{c}
  {\rule[-3mm]{0mm}{8mm}$F_{\rm M}(1{\rm S}_{1/2},Z\alpha)$} &
\multicolumn{2}{c}
  {\rule[-3mm]{0mm}{8mm}$F_{\rm M}(2{\rm S}_{1/2},Z\alpha)$} &
\multicolumn{2}{c}
  {\rule[-3mm]{0mm}{8mm}$F_{\rm M}(2{\rm P}_{1/2},Z\alpha)$} &
\multicolumn{2}{c}
  {\rule[-3mm]{0mm}{8mm}$F_{\rm M}(2{\rm P}_{3/2},Z\alpha)$}
\\
\hline
$1$  &
 $4$ & $385~373~827~7(1)$ &
 $4$ & $505~635~023(2)$ &
 $0$ & $854~995~085(2)$ &
 $0$ & $607~432~559(2)$ \\
$2$  &
 $4$ & $286~403~675~7(1)$ &
 $4$ & $407~404~458(2)$ &
 $0$ & $854~987~226(2)$ &
 $0$ & $607~237~953(2)$ \\
$3$  &
 $4$ & $195~185~543~6(1)$ &
 $4$ & $317~349~564(2)$ &
 $0$ & $854~984~972(2)$ &
 $0$ & $606~925~899(2)$ \\
$4$  &
 $4$ & $110~233~778~8(1)$ &
 $4$ & $233~940~436(2)$ &
 $0$ & $854~995~198(2)$ &
 $0$ & $606~504~344(2)$ \\
$5$  &
 $4$ & $030~578~931~1(1)$ &
 $4$ & $156~170~778(2)$ &
 $0$ & $855~023~557(2)$ &
 $0$ & $605~979~905(2)$ \\
\end{tabular}
\end{center}
\end{minipage}
\end{center}
\end{table}

\narrowtext

\typeout{Subsection 3C}
%
%
\subsection{The Middle-Energy Remainder}
\label{MiddleEnergyRemainder}

The remainder term in the middle-energy part involves the propagator
$G_{\rm B}$ defined in Eq.~(\ref{DefinitionOfGB}), $G_{\rm B} = G -
G_{\rm A}$, where $G$ is defined in (\ref{DefinitionOfG}) and $G_{\rm
A}$ is given in (\ref{DefinitionOfGA}).  In analogy to the
middle-energy subtraction term, the middle-energy remainder can be
rewritten as a three-dimensional integral,
\begin{eqnarray}
\label{EvaluationOfEMB}
\Delta E_{\rm MB} &=& \frac{\alpha}{\pi} \, 
  \int_0^{\lfrac{9}{10}\,E_n} {\rm d}z 
\nonumber\\
& & \quad \int_0^1 {\rm d}r \, \int_0^\infty {\rm d}y \,
S_{\rm MB}(r,y,z)\,,
\end{eqnarray}
where
\begin{eqnarray}
\label{defSMB}
& & S_{\rm MB}(r,y,z) = - \frac{2 \, r^2 \, y^5}{a^6} \, 
\sum_{|\kappa|=1}^{\infty} \sum_{\kappa = \pm |\kappa|} 
\sum_{i,j=1}^{2} \, f_{\bar \imath}\left(\frac{r\,y}{a}\right)
\nonumber\\
& & \quad \times \,\,
G^{ij}_{{\rm B},\kappa}\left(\frac{r\,y}{a},\frac{y}{a},z\right) \, 
f_{\bar \jmath}\left(\frac{y}{a}\right) \, 
A^{ij}_{\kappa}\left(\frac{r\,y}{a},\frac{y}{a}\right)\,.
\end{eqnarray}
The functions $G^{ij}_{{\rm B},\kappa}$ are obtained as the difference
of the expansion of the full propagator $G$ and the simplified
propagator $G_{\rm A}$ into angular momenta,
\begin{equation}
G^{ij}_{{\rm B},\kappa} = G^{ij}_{\kappa} -
  G^{ij}_{{\rm A},\kappa} \,,
\end{equation}
where the $G^{ij}_{\kappa}$ are listed in Eq.~(A.16) in~\cite{Mo1974a}
and in Eq.~(D.43) in~\cite{Je1999}, and the $G^{ij}_{{\rm A},\kappa}$
have already been defined in Eqs.~(5.4) and (A.20) in \cite{Mo1974a} and in
Eqs.~(D.37) and (D.42) in~\cite{Je1999}.  There are no lower-order
terms to subtract, and therefore
\begin{equation}
\label{DefinitionOfFMB}
\Delta E_{\rm MB} = \frac{\alpha}{\pi} \,
\frac{(Z\alpha)^4}{n^3}\,F_{\rm MB}(nl_j,Z\alpha)\,.
\end{equation}
The three-dimensional integral (\ref{EvaluationOfEMB}) is evaluated by
successive Gaussian quadrature. Details of the integration procedure
are provided in~\cite{Je1999}. Numerical results for the middle-energy
remainder $F_{\rm MB}$ are summarized in Table~\ref{tableFMKL} for
the K- and L-shell states.

For the middle-energy part, the separation into a subtraction and a
remainder term has considerable computational advantages which become
obvious upon inspection of Eqs.~(\ref{DefinitionOfFMA})
and~(\ref{DefinitionOfFMB}).  The subtraction involves a propagator
whose angular components can be evaluated by 
recursion~\cite{Mo1974b,Je1999}, which is computationally
time-consuming. Because the subtraction term involves lower-order components
[see Eq.~(\ref{DefinitionOfFM})], it has to be evaluated to high
precision numerically (in a typical case, a relative uncertainty of
$10^{-19}$ is required).  This high precision requires in turn a large
number of integration points for the Gaussian quadratures, which is
possible only if the numerical evaluation of the integrand is not
computationally time-consuming.  For the remainder term, no lower-order
terms have to be subtracted, and the relative precision required of the
integrals is in the range of $10^{-11}\dots10^{-9}$.  A numerical
evaluation to this lower level of precision is feasible, although the
calculation of the Green function $G_{\rm B}$ is computationally more
time consuming than that of $G_{\rm A}$~\cite{Mo1974a,Mo1974b,Je1999}.  The
separation of the high-energy part into a subtraction term and a
remainder term, which is discussed in Sec.~\ref{HighEnergyPart}, is
motivated by analogous considerations as for the middle-energy part.
In the high-energy part, this separation is even more important than
in the middle-energy part, because of the occurrence of infinite terms
which need to be subtracted analytically before a numerical evaluation
can proceed [see Eq.~(\ref{DeltaEHArecovered}) below].

We now summarize the results for the middle-energy part.
The middle-energy part is the sum of the middle-energy subtraction
term $F_{\rm MA}$ and the middle-energy remainder $F_{\rm MB}$ [see
also Eq.~(\ref{DecompositionOfFM})]. Numerical results are summarized
in Table~\ref{tableFMKL} for the K- and L-shell states. The low-energy
part $F_{\rm L}$ is defined as the sum of the infrared contribution
$F_{\rm IR}$ and the middle-energy contribution $F_{\rm M}$ [see
Eq.~(\ref{FLasSum})]. The results for $F_{\rm L}$ are provided in the
Table~\ref{tableFLKL} for the K- and L-shell states.  The 
limits for the low-energy part as a function of the bound state
quantum numbers can be found in Eq.~(7.80) of~\cite{Je1999}:
\begin{eqnarray}
\lefteqn{F_{\rm L}(nl_j,Z\alpha) =
  \frac{4}{3}\,\delta_{l,0}\,\ln(Z\alpha)^{-2} } \nonumber\\[2ex]
& & \quad - \frac{4}{3}\,\ln k_0(n,l)
+ \left(\ln2 - \frac{11}{10}\right) \, \frac{1}{n} \nonumber\\
& & + \left(2\,\ln2 - \frac{16}{15}\right) \, \frac{1}{2\,l + 1}
+ \left(\frac{3}{2}\,\ln2 - \frac{7}{4}\right)\,
\frac{1}{\kappa\,(2\,l+1)}
\nonumber \\
& & \quad +
\left(-\frac{3}{2}\,\ln2 + \frac{9}{4}\right)\,\frac{1}{|\kappa|}
+ \left(\frac{4}{3}\,\ln2 - \frac{1}{3}\right) \delta_{l,0}
\nonumber \\
& & \quad +
\left(\ln2 - \frac{5}{6}\right) \frac{n-2\,l-1}{n\,(2\,l+1)} +
{\rm O}(Z\alpha)\,.
\end{eqnarray}
The limits for the
states under investigation in this paper are
\begin{eqnarray}
\label{AsympFL}
F_{\rm L}(1{\rm S}_{1/2},Z \alpha) &=& 
 (4/3) \, \ln (Z\alpha)^{-2} - 1.554~642 + {\rm O}(Z\alpha) \,,
\nonumber\\[2ex]
F_{\rm L}(2{\rm S}_{1/2},Z \alpha) &=&  
 (4/3) \, \ln (Z\alpha)^{-2} - 1.191~497 + {\rm O}(Z\alpha) \,,
\nonumber\\[2ex] 
F_{\rm L}(2{\rm P}_{1/2},Z \alpha) &=&
 0.940~023 + {\rm O}(Z\alpha) \,,
\nonumber\\[2ex] 
F_{\rm L}(2{\rm P}_{3/2},Z \alpha) &=&
 0.690~023 + {\rm O}(Z\alpha) \,.
\end{eqnarray}
These limits are consistent with the numerical data in
Table~\ref{tableFLKL}. For S states, the
low-energy contribution $F_{\rm L}$ diverges logarithmically as
$Z\alpha\to0$, whereas for P states, $F_{\rm L}$ approaches a constant
as $Z\alpha\to0$. The leading logarithm is a consequence of an
infrared divergence cut off by the atomic momentum scale. It is a
nonrelativistic effect which is generated by the nonvanishing
probability density of S waves at the origin in the 
nonrelativistic limit. The presence of the logarithmic behavior
for S states [nonvanishing $A_{41}$-coefficient, 
see Eqs.~(\ref{defFLO}) and (\ref{defFLOnS})] and its
absence for P states
is reproduced consistently by the data in~Table~\ref{tableFLKL}.

\widetext

%
%
\begin{table}[thb]
\begin{center}
\begin{minipage}{14cm}
\begin{center}
\caption{\label{tableFLKL} Low-energy
part $F_{\rm L}$ for the K- and L-shell states
$F_{\rm L}(1{\rm S}_{1/2},Z\alpha)$,
$F_{\rm L}(2{\rm S}_{1/2},Z\alpha)$,
$F_{\rm L}(2{\rm P}_{1/2},Z\alpha)$,
and $F_{\rm L}(2{\rm P}_{3/2},Z\alpha)$,
evaluated for low-$Z$
hydrogenlike ions.}
\begin{tabular}{lr@{.}lr@{.}lr@{.}lr@{.}l}
\multicolumn{1}{c}{$Z$} &
\multicolumn{2}{c}
  {\rule[-3mm]{0mm}{8mm}$F_{\rm L}(1{\rm S}_{1/2},Z\alpha)$} &
\multicolumn{2}{c}
  {\rule[-3mm]{0mm}{8mm}$F_{\rm L}(2{\rm S}_{1/2},Z\alpha)$} &
\multicolumn{2}{c}
  {\rule[-3mm]{0mm}{8mm}$F_{\rm L}(2{\rm P}_{1/2},Z\alpha)$} &
\multicolumn{2}{c}
  {\rule[-3mm]{0mm}{8mm}$F_{\rm L}(2{\rm P}_{3/2},Z\alpha)$}
\\
\hline
$1$  &
 $11$ & $621~997~564~5(1)$ &
 $11$ & $985~399~203(2)$ &
 $ 0$ & $940~322~937(2)$ &
 $ 0$ & $690~169~056(2)$ \\
$2$  &
  $9$ & $825~405~794~7(1)$ &
 $10$ & $189~430~095(2)$ &
 $ 0$ & $941~060~895(2)$ &
 $ 0$ & $690~517~414(2)$ \\
$3$  &
  $8$ & $793~341~365~4(1)$ &
 $ 9$ & $158~273~526(2)$ &
 $ 0$ & $942~147~482(2)$ &
 $ 0$ & $691~017~729(2)$ \\
$4$  &
  $8$ & $073~357~919~4(1)$ &
 $ 8$ & $439~442~234(2)$ &
 $ 0$ & $943~538~386(2)$ &
 $ 0$ & $691~645~132(2)$ \\
$5$  &
  $7$ & $523~832~250~6(1)$ &
 $ 7$ & $891~285~736(2)$ &
 $ 0$ & $945~204~392(2)$ &
 $ 0$ & $692~383~083(2)$ \\
\end{tabular}
\end{center}
\end{minipage}
\end{center}
\end{table}

\narrowtext

\typeout{Section 4}
%
%
\section{The High-Energy Part}
\label{HighEnergyPart}

\typeout{Subsection 4A}
%
%
\subsection{The High-Energy Subtraction Term}
\label{HighEnergySubtraction}

The high-energy part is given by
\begin{eqnarray}
\label{deltaEH}
\lefteqn{\Delta E_{\rm H} = 
- \lim_{\Lambda\to\infty}
{\rm i}\, e^2 \,
\int_{C_{\rm H}} \! \frac{d\omega}{2 \pi}
\int \! \! \frac{d^3 \bbox{k}}{(2 \pi)^3} \,
D_{\mu\nu}(k^2,\Lambda)} \nonumber\\[2ex]
& & \times \left< \psi \left| \, \alpha^{\mu} \,
{\rm e}^{{\rm i}\,\bbox{k}\cdot\bbox{x}}\,
G(E_n - \omega) \, \alpha^{\nu} \, 
{\rm e}^{-{\rm i}\,\bbox{k}\cdot\bbox{x}}\,
\right| \psi \right> \,,
\end{eqnarray}
where relevant definitions of the symbols can be found in
Eqs.~(\ref{deltaESEM})--(\ref{PhotonPropagator}), and
the contour $C_{\rm H}$ is as shown in Fig.~\ref{IntegrationContour}.
The high-energy part consists of the following integration region for
the virtual photon,
\begin{equation}
\left.
\begin{array}{l}
\omega \in 
\left(E_n-{\rm i}\,\infty,E_n+{\rm i}\,\infty\right) \\[1ex]  
z \in \left(-{\rm i}\,\infty,{\rm i}\,\infty\right) 
\end{array}
\right\}
\mbox{high-energy part $\Delta E_{\rm H}$}\,.
\end{equation}
The separation of the high-energy part into a subtraction term and a
remainder is accomplished as in the middle-energy part [see
Eq.~(\ref{DecompositionOfG})] by writing the full Dirac-Coulomb Green
function $G$ [Eq.~(\ref{DefinitionOfG})] as $G = G_{\rm A} + G_{\rm
B}$.  We define the high-energy subtraction term $F_{\rm HA}$ as the
expression obtained upon substitution of the propagator $G_{\rm A}$
for $G$ in Eq.~(\ref{deltaEH}), and a substitution of the propagator
$G_{\rm B}$ for $G$ in Eq.~(\ref{deltaEH}) leads to the high-energy
remainder $F_{\rm HB}$ which is discussed in
Sec.~\ref{HighEnergyRemainder}.  The subtraction term (including all
divergent contributions) is generated by $G_{\rm A}$, the high-energy
remainder term corresponds to $G_{\rm B}$.  We have
\begin{eqnarray}
\label{deltaEHA}
\lefteqn{\Delta E_{\rm HA} =
- \lim_{\Lambda\to\infty}
{\rm i}\, e^2 \,
\int_{C_{\rm H}} \! \frac{d\omega}{2 \pi}
\int \! \! \frac{d^3 \bbox{k}}{(2 \pi)^3} \,
D_{\mu\nu}(k^2,\Lambda)} \nonumber\\[2ex]
& &  \times \left< \psi \left| \, \alpha^{\mu} \,
{\rm e}^{{\rm i}\,\bbox{k}\cdot\bbox{x}}\,
G_{\rm A}(E_n - \omega) \, \alpha^{\nu} \, 
{\rm e}^{-{\rm i}\,\bbox{k}\cdot\bbox{x}}\,
\right| \psi \right> 
\end{eqnarray}
and
\begin{eqnarray}
\label{deltaEHB}
\lefteqn{\Delta E_{\rm HB} =
{\rm -i}\, e^2 \,
\int_{C_{\rm H}} \! \frac{d\omega}{2 \pi}
\int \! \! \frac{d^3 \bbox{k}}{(2 \pi)^3} \,
D_{\mu\nu}(k^2)} \nonumber\\[1ex]
& & \times \left< \psi \left| \, \alpha^{\mu} \,
{\rm e}^{{\rm i}\,\bbox{k}\cdot\bbox{x}}\,
G_{\rm B}(E_n - \omega) \, \alpha^{\nu} \, 
{\rm e}^{-{\rm i}\,\bbox{k}\cdot\bbox{x}}\,
\right| \psi \right> \,.
\end{eqnarray}
The contribution $\Delta E_{\rm HA}$ corresponding to $G_{\rm A}$ can
be separated further into a term $\Delta E^{(1)}_{\rm HA}$, which
contains all divergent contributions, and a term $\Delta E^{(2)}_{\rm
HA}$, which contains contributions of lower order than $(Z\alpha)^4$,
but is convergent as $\Lambda \to \infty$.  This separation is
described in detail in~\cite{Mo1974a,Mo1973}.  We have
\begin{equation}
\Delta E_{\rm HA} = \Delta E^{(1)}_{\rm HA} +
\Delta E^{(2)}_{\rm HA}\,.
\end{equation}
We obtain for $\Delta E^{(1)}_{\rm HA}$, which contains a 
logarithmic divergence as $\Lambda \to \infty$,
\begin{eqnarray}
\Delta E^{(1)}_{\rm HA} &=& \frac{\alpha}{\pi} \,
\left[ \left(\frac{3}{4}\,\ln \Lambda^2 - \frac{9}{8} \right) \,
\langle \beta \rangle +
\left(\frac{1}{2}\,\ln2 - \frac{17}{12} \right) \, \langle V \rangle 
\right. \nonumber\\[2ex]
& & \qquad \left. + \frac{(Z\alpha)^4}{n^3} \, 
F^{(1)}_{\rm HA}(nl_j,Z\alpha) \right]\,.
\end{eqnarray}
For the contribution $F^{(1)}_{\rm HA}$, an explicit analytic result
is given in Eq.~(4.15) in~\cite{Mo1974a}. This contribution is
therefore not discussed in any further detail here. The contribution
$\Delta E^{(2)}_{\rm HA}$ contains lower-order terms:
\begin{eqnarray}
\Delta E^{(2)}_{\rm HA} &=& \frac{\alpha}{\pi} \,
\left[\left(-\frac{1}{2}\,\ln2 + \frac{1}{4} \right) \, \langle V \rangle
\right. \nonumber\\[1ex]
& & \quad\quad\quad\quad \left. + \frac{(Z\alpha)^4}{n^3} \,
F^{(2)}_{\rm HA}(nl_j,Z\alpha) \right]\,.
\end{eqnarray}
Altogether we have
\begin{eqnarray}
\label{DeltaEHArecovered}
\Delta E_{\rm HA} &=&
\Delta E^{(1)}_{\rm HA} + \Delta E^{(2)}_{\rm HA}
\nonumber\\[2ex]
&=& \frac{\alpha}{\pi} \,
\left[ \left(\frac{3}{4}\,\ln \Lambda^2 - \frac{9}{8} \right) \,
\langle \beta \rangle -
\frac{7}{6} \, \langle V \rangle
\right.
\nonumber\\[1ex]
& & \quad\quad \left. + \frac{(Z\alpha)^4}{n^3} \,
F_{\rm HA}(nl_j,Z \alpha) \right]\,.
\end{eqnarray}
The scaled function $F_{\rm HA}(nl_j,Z \alpha)$ is given by
\begin{equation}
F_{\rm HA}(nl_j,Z \alpha) =
F^{(1)}_{\rm HA}(nl_j,Z \alpha) +
F^{(2)}_{\rm HA}(nl_j,Z \alpha)\,.
\end{equation}
The term $\Delta
E^{(2)}_{\rm HA}$ falls naturally into a sum of
four contributions~\cite{Mo1974a},
\begin{equation}
\Delta E^{(2)}_{\rm HA} = T_1 + T_2 + T_3 + T_4 \,,
\end{equation}
where
\begin{eqnarray}
T_1 &=& -\frac{1}{10} \langle V \rangle + \frac{(Z\alpha)^4}{n^3} \, 
h_1(nl_j,Z \alpha)\,, \nonumber\\
T_2 &=& \left(\frac{7}{20} - \frac{1}{2} \, \ln2 \right)
\, \langle V \rangle + 
\frac{(Z\alpha)^4}{n^3} \, h_2(nl_j,Z \alpha)\,, \nonumber\\
T_3 &=& \frac{(Z\alpha)^4}{n^3} \, h_3(nl_j,Z \alpha)\,, \nonumber\\
T_4 &=& \frac{(Z\alpha)^4}{n^3} \, h_4(nl_j,Z \alpha)\,.
\end{eqnarray}
The functions $h_i$ ($i=1,2,3,4$) are defined in Eqs.~(4.18), 
(4.19) and (4.21) in~\cite{Mo1974a} (see also Eq.~(3.6) in~\cite{Mo1982}).
The evaluation of the high-energy subtraction term proceeds as outlined
in~\cite{Mo1974a,Mo1974b,Mo1982}, albeit with an increased accuracy
and improved calculational methods 
in intermediate steps of the calculation in order to overcome the
severe numerical cancellations in the low-$Z$ region.
We recover $F^{(2)}_{\rm HA}$ as the sum
\begin{eqnarray}
& & F^{(2)}_{\rm HA}(nl_j,Z \alpha) = 
h_1(nl_j,Z\alpha) + h_2(nl_j,Z\alpha) 
\nonumber\\[2ex]
& & \quad + h_3(nl_j,Z\alpha) + h_4(nl_j,Z\alpha)\,.
\end{eqnarray}
The scaled function $F_{\rm HA}(nl_j,Z \alpha)$ [see also
Eqs.~(\ref{DefinitionOfFH}) and (\ref{DecompositionOfFH})] is given by
\begin{equation}
F_{\rm HA}(nl_j,Z \alpha) = 
F^{(1)}_{\rm HA}(nl_j,Z \alpha) + 
F^{(2)}_{\rm HA}(nl_j,Z \alpha)\,.
\end{equation}
The limits of the contributions $F^{(1)}_{\rm HA}(nl_j,Z \alpha)$
and $F^{(2)}_{\rm HA}(nl_j,Z\alpha)$ as $(Z\alpha)\to 0$ have been
investigated in~\cite{Mo1974a,Mo1973,Mo1982}.  For the contribution
$F^{(1)}_{\rm HA}(nl_j,0)$, the result can be found in Eq.~(3.5)
in~\cite{Mo1982}.  The limits of the functions
$h_i(nl_j,Z\alpha)$ ($i=1,2,3,4$) as $Z \alpha \to 0$ are given as a
function of the atomic state quantum numbers in Eq.~(3.8)
in~\cite{Mo1982}.  For the scaled high-energy subtraction term $F_{\rm
HA}$, the limits read (see Eq.~(3.9) in~\cite{Mo1982})
\begin{eqnarray}
\label{limFHAgen}
\lefteqn{F_{\rm HA}(nl_j,Z\alpha) = 
\left(\frac{11}{10} - \ln2\right) \, \frac{1}{n}}  \nonumber\\
& & \;\;\;\;\;
+ \left(\frac{16}{15} - 2 \, \ln2\right) \, \frac{1}{2\,l + 1}
+ \left(\frac{1}{2}\,\ln2 - \frac{1}{4}\right) 
\frac{1}{\kappa\,(2\,l+1)}
\nonumber\\
& & \;\;\;\;\;
+ \left(\frac{3}{2}\,\ln2 - \frac{9}{4}\right)\,\frac{1}{|\kappa|}
+ {\rm O}(Z\alpha)\,.
\end{eqnarray}
Therefore, the explicit forms of the 
limits for the states under investigation in this
paper are
\begin{eqnarray}
\label{limFHA}
F_{\rm HA}(1{\rm S}_{1/2},Z \alpha) &=& 
-1.219~627 + {\rm O}(Z\alpha)\,, \nonumber\\[2ex]
F_{\rm HA}(2{\rm S}_{1/2},Z \alpha) &=&
-1.423~054 + {\rm O}(Z\alpha)\,, \nonumber\\[2ex]
F_{\rm HA}(2{\rm P}_{1/2},Z \alpha) &=&
-1.081~204 + {\rm O}(Z\alpha)\,, \nonumber\\[2ex]
F_{\rm HA}(2{\rm P}_{3/2},Z \alpha) &=&
-0.524~351 + {\rm O}(Z\alpha)\,. 
\end{eqnarray}
Numerical results for $F_{\rm HA}$, which are presented in
Table~\ref{tableFHKL}, exhibit consistency with the limits in
Eq.~(\ref{limFHA}).

\widetext

%
%
\begin{table}[htb]
\begin{center}
\begin{minipage}{17cm}
\begin{center}
\caption{\label{tableFHKL} Numerical results for the
high-energy subtraction term $F_{\rm HA}$
and the high-energy remainder term $F_{\rm HB}$.
The high-energy term $F_{\rm H}$ is the sum 
$F_{\rm H}(nl_j,Z\alpha) = 
F_{\rm HA}(nl_j,Z\alpha) + F_{\rm HB}(nl_j,Z\alpha)$.}
\begin{tabular}{lr@{.}lr@{.}lr@{.}lr@{.}l}
$Z$ & 
\multicolumn{2}{c}{\rule[-3mm]{0mm}{8mm} 
  $F_{\rm HA}(1{\rm S}_{1/2},Z\alpha)$} &
\multicolumn{2}{c}{\rule[-3mm]{0mm}{8mm} 
  $F_{\rm HA}(2{\rm S}_{1/2},Z\alpha)$} &
\multicolumn{2}{c}{\rule[-3mm]{0mm}{8mm} 
  $F_{\rm HA}(2{\rm P}_{1/2},Z\alpha)$} &
\multicolumn{2}{c}{\rule[-3mm]{0mm}{8mm} 
  $F_{\rm HA}(2{\rm P}_{3/2},Z\alpha)$} \\
\hline
$1$ &
 $-1$ & $216~846~660~6(1)$ & 
 $-1$ & $420~293~291(1)$ &
 $-1$ & $081~265~954(1)$ &
 $-0$ & $524~359~802(1)$ \\
$2$ &
 $-1$ & $214~322~536~9(1)$ & 
 $-1$ & $417~829~864(1)$ &
 $-1$ & $081~451~269(1)$ &
 $-0$ & $524~385~053(1)$ \\
$3$ &
 $-1$ & $212~026~714~1(1)$ & 
 $-1$ & $415~635~310(1)$ &
 $-1$ & $081~760~224(1)$ &
 $-0$ & $524~427~051(1)$ \\
$4$ &
 $-1$ & $209~942~847~4(1)$ &
 $-1$ & $413~693~422(1)$ &
 $-1$ & $082~192~995(1)$ &
 $-0$ & $524~485~727(1)$ \\
$5$ &
 $-1$ & $208~059~033~6(1)$ &
 $-1$ & $411~992~480(1)$ &
 $-1$ & $082~749~845(1)$ &
 $-0$ & $524~561~017(1)$ \\
\hline
\hline
$Z$ & 
\multicolumn{2}{c}{\rule[-3mm]{0mm}{8mm} 
  $F_{\rm HB}(1{\rm S}_{1/2},Z\alpha)$} &
\multicolumn{2}{c}{\rule[-3mm]{0mm}{8mm} 
  $F_{\rm HB}(2{\rm S}_{1/2},Z\alpha)$} &
\multicolumn{2}{c}{\rule[-3mm]{0mm}{8mm} 
  $F_{\rm HB}(2{\rm P}_{1/2},Z\alpha)$} &
\multicolumn{2}{c}{\rule[-3mm]{0mm}{8mm} 
  $F_{\rm HB}(2{\rm P}_{3/2},Z\alpha)$} \\
\hline
$1$ &
 $-0$ & $088~357~254(1)$ &
 $-0$ & $018~280~727(5)$\tablenote{Result obtained with a 
greater number of integration nodes than are used for the higher-$Z$
results.} &
 $ 0$ & $014~546~64(1)$ &
 $-0$ & $042~310~69(1)$ \\
$2$ &
 $-0$ & $082~758~206(1)$ &
 $-0$ & $012~729~99(1)$ &
 $ 0$ & $014~574~21(1)$ &
 $-0$ & $042~296~81(1)$ \\
$3$ &
 $-0$ & $076~811~229(1)$ &
 $-0$ & $006~861~02(1)$ &
 $ 0$ & $014~620~51(1)$ &
 $-0$ & $042~273~58(1)$ \\
$4$ &
 $-0$ & $070~590~991(1)$ &
 $-0$ & $000~746~40(1)$ &
 $ 0$ & $014~685~82(1)$ &
 $-0$ & $042~240~92(1)$ \\
$5$ &
 $-0$ & $064~146~139(1)$ &
 $ 0$ & $005~567~16(1)$ &
 $ 0$ & $014~770~52(1)$ &
 $-0$ & $042~198~76(1)$ \\
\hline
\hline
$Z$ & 
\multicolumn{2}{c}{\rule[-3mm]{0mm}{8mm} 
  $F_{\rm H}(1{\rm S}_{1/2},Z\alpha)$} &
\multicolumn{2}{c}{\rule[-3mm]{0mm}{8mm} 
  $F_{\rm H}(2{\rm S}_{1/2},Z\alpha)$} &
\multicolumn{2}{c}{\rule[-3mm]{0mm}{8mm} 
  $F_{\rm H}(2{\rm P}_{1/2},Z\alpha)$} &
\multicolumn{2}{c}{\rule[-3mm]{0mm}{8mm} 
  $F_{\rm H}(2{\rm P}_{3/2},Z\alpha)$} \\
\hline
$1$ &
 $-1$ & $305~203~915(1)$ &
 $-1$ & $438~574~018(5)$ &
 $-1$ & $066~719~31(1)$ &
 $-0$ & $566~670~50(1)$ \\
$2$ &
 $-1$ & $297~080~743(1)$ &
 $-1$ & $430~559~85(1)$ &
 $-1$ & $066~877~06(1)$ &
 $-0$ & $566~681~86(1)$ \\
$3$ &
 $-1$ & $288~837~943(1)$ &
 $-1$ & $422~496~33(1)$ &
 $-1$ & $067~139~72(1)$ &
 $-0$ & $566~700~63(1)$ \\
$4$ &
 $-1$ & $280~533~839(1)$ &
 $-1$ & $414~439~82(1)$ &
 $-1$ & $067~507~18(1)$ &
 $-0$ & $566~726~65(1)$ \\
$5$ &
 $-1$ & $272~205~173(1)$ &
 $-1$ & $406~425~32(1)$ &
 $-1$ & $067~979~33(1)$ &
 $-0$ & $566~759~78(1)$ \\
\end{tabular}
\end{center}
\end{minipage}
\end{center}
\end{table}

\narrowtext

\typeout{Subsection 4B}
%
%
\subsection{The High-Energy Remainder}
\label{HighEnergyRemainder}

The remainder term in the high-energy part involves the
propagator $G_{\rm B}$ defined in Eq.~(\ref{DefinitionOfGB}),
$G_{\rm B} = G - G_{\rm A}$,
where $G$ is defined in (\ref{DefinitionOfG}) and $G_{\rm A}$ is given
in (\ref{DefinitionOfGA}).  The energy shift is
\begin{eqnarray}
\label{IntegralEHB}
\Delta E_{\rm HB} &=& - \frac{{\rm i}\,\alpha}{\pi} \, 
\int_{0}^{i\,\infty} \! {\rm d}z \,
\int_0^\infty \! {\rm d}x_1 \, x_1^2 
\nonumber\\[1ex]
& & \quad \,
\int_0^{x_1} \! {\rm d}x_2 \, x_2^2 \,\, 
\left\{{\cal M}_{\rm HB}(x_2,x_1,z) + {\rm c.c.}\right\}\,,
\end{eqnarray}
where c.c. denotes the complex conjugate. The photon energy integration
is evaluated with the aid of the substitution
\begin{equation}
\label{SubstitutionHB}
z \to {\rm i}\,u \;\;\;\;\;\;\; \mbox{where} \;\;\;\;\;\;\;
u = \frac{1}{2}\,\left(\frac{1}{t} - t\right)\,.
\end{equation}
In analogy with the middle-energy subtraction and remainder 
terms discussed in Secs.~\ref{MiddleEnergySubtraction} 
and~\ref{MiddleEnergyRemainder} [see especially 
Eqs.~(\ref{defSMA}) and (\ref{defSMB})], 
the functions ${\cal M}_{\rm HB}(x_2,x_1,z)$ and
$S_{\rm HB}(r,y,z)$ and the terms $T_{{\rm HB},|\kappa|}$
are defined implicitly in the following:
\begin{eqnarray}
\label{defSHB}
\lefteqn{S_{\rm HB}(r,y,t) =} \nonumber\\[2ex] 
&=& \left(1 + \frac{1}{t^2}\right)\,
\frac{r^2\,y^5}{a^6} \, {\rm Re}\left[
{\cal M}_{\rm HB}\left(\frac{r\,y}{a},\frac{y}{a},
  {\rm i}\,u\right)
\right] \nonumber\\[2ex]
&=& \left(1 + \frac{1}{t^2}\right)\,
\frac{r^2\,y^5}{a^6} \,
\sum_{|\kappa|=1}^{\infty} \,\, \sum_{\kappa = \pm |\kappa|} \,\,
\sum_{i,j=1}^{2} \nonumber\\[1ex]
& &  
{\rm Re}\left[
f_i\left(\frac{r\,y}{a}\right) \,
G^{ij}_{{\rm B},\kappa}\left(\frac{r\,y}{a},\frac{y}{a},
 {\rm i}\,u \right) \,
f_j\left(\frac{y}{a}\right)\,
{\cal A}_{\kappa}\left(\frac{r\,y}{a},\frac{y}{a}\right)\right. 
\nonumber\\[1ex]
& & 
\left. - f_{\bar \imath}\left(\frac{r\,y}{a}\right) \,
G^{ij}_{{\rm B},\kappa}\left(\frac{r\,y}{a},\frac{y}{a},
{\rm i}\,u \right)\,
f_{\bar \jmath}\left(\frac{y}{a}\right) \, 
{\cal A}^{ij}_{\kappa}\left(\frac{r\,y}{a},\frac{y}{a}\right)
\right] \nonumber\\[2ex]
&=& \left(1 + \frac{1}{t^2}\right)\,
\frac{r^2\,y^5}{a^6}\, 
\sum_{|\kappa|=1}^{\infty} 
T_{{\rm HB},|\kappa|}(r,y,t)\,.
\end{eqnarray}
The only substantial 
difference from the treatment of the middle-energy remainder
lies in the prefactor generated by the parameterization of the
complex photon energy given in Eq.~(\ref{SubstitutionHB}).
The photon angular functions ${\cal A}_{\kappa}$ and ${\cal
A}^{ij}_{\kappa}$ ($i,j=1,2$) {\em for the high-energy part} are
defined in Eq.~(5.8) of Ref.~\cite{Mo1974a} and 
in Eq.~(4.3) in~\cite{Mo1982} for an arbitrary bound
state. Special formulas for the 
ground state can be found in Eq.~(5.9) 
of Ref.~\cite{Mo1974a}. The functions ${\cal A}_{\kappa}$ and 
${\cal A}^{ij}_{\kappa}$
are {\em not} identical to the photon angular functions for the
infrared and middle-energy parts $A^{ij}_{\kappa}$ ($i,j=1,2$) which
are used for the low-energy part of the calculation in
Sec.~\ref{LowEnergyPart}.
It might be worth mentioning that
in~\cite{Mo1974a,Mo1974b,Mo1982,Mo1992},
both the functions $A^{ij}_{\kappa}$ and ${\cal
A}^{ij}_{\kappa}$ are denoted by the symbol $A^{ij}_{\kappa}$. It is
clear from the context which of the functions is employed in each case.

In the last line
of Eq.~(\ref{defSHB}), we implicitly define the terms
$T_{{\rm HB},|\kappa|}$ as
\begin{eqnarray}
\label{DefinitionOfTHBkappa}
\lefteqn{T_{{\rm HB},|\kappa|}(r,y,t) =
\sum_{\kappa = \pm |\kappa|} \,\,
\sum_{i,j=1}^{2}} \nonumber\\[1ex]
& & \,\, {\rm Re} \left[
f_i\left(\frac{r\,y}{a}\right) \,
G^{ij}_{{\rm B},\kappa}\left(\frac{r\,y}{a},\frac{y}{a},
  {\rm i}\,u\right) \,
f_j\left(\frac{y}{a}\right)\right.\nonumber\\[1ex]
& & \quad \times \left.
{\cal A}_{\kappa}\left(\frac{r\,y}{a},\frac{y}{a}\right) 
\right. \nonumber\\[2ex]
& & \left. - f_{\bar \imath}\left(\frac{r\,y}{a}\right) \,
G^{ij}_{{\rm B},\kappa}\left(\frac{r\,y}{a},\frac{y}{a},
  {\rm i}\,u\right)\,
f_{\bar \jmath}\left(\frac{y}{a}\right) \right. \nonumber\\[1ex]
& & \quad \times \left.
{\cal A}^{ij}_{\kappa}\left(\frac{r\,y}{a},\frac{y}{a}\right)\right].
\end{eqnarray}
With these definitions, the high-energy remainder can be rewritten as
\begin{equation}
\label{EvaluationOfEHB}
\Delta E_{\rm HB} = \frac{\alpha}{\pi} \, 
\int_0^1 {\rm d}t \, \int_0^1 {\rm d}r \, \int_0^\infty {\rm d}y \, 
S_{\rm HB}(r,y,t)\,.
\end{equation}
There are no lower-order terms to subtract, and therefore
\begin{equation}
\label{DefinitionOfFHB}
\Delta E_{\rm HB} = \frac{\alpha}{\pi} \,
\frac{(Z\alpha)^4}{n^3}\,F_{\rm HB}(nl_j,Z\alpha)\,.
\end{equation}
For the high-energy remainder $F_{\rm HB}$, the limits as
$Z\alpha\to0$ read [see Eq.~(4.15) in \cite{Mo1982}]
\begin{eqnarray}
\label{limFHBgen}
\lefteqn{F_{\rm HB}(nl_j,Z\alpha) =
\frac{1}{2\,l+1} \,
\left[\left(\frac{17}{18} - \frac{4}{3}\,\ln2\right) \delta_{l,0} \right.}
\nonumber\\
&&
\quad \left. +
\left(\frac{3}{2} - 2\,\ln2\right) \frac{1}{\kappa} \right.
\nonumber\\
&&
\quad \left. +
\left(\frac{5}{6} - \ln2\right) \frac{n-2\,l-1}{n}\right] +
{\rm O}(Z\alpha)\,.
\end{eqnarray}
For the atomic states under investigation, this leads to
\begin{eqnarray}
\label{limFHB}
F_{\rm HB}(1{\rm S}_{1/2},Z\alpha) &=& 
-0.093~457 + {\rm O}(Z\alpha) \,,\nonumber\\[2ex]
F_{\rm HB}(2{\rm S}_{1/2},Z\alpha) &=&
-0.023~364 + {\rm O}(Z\alpha) \,,\nonumber\\[2ex]
F_{\rm HB}(2{\rm P}_{1/2},Z\alpha) &=&
0.014~538 + {\rm O}(Z\alpha) \,,\nonumber\\[2ex]
F_{\rm HB}(2{\rm P}_{3/2},Z\alpha) &=&
-0.042~315 + {\rm O}(Z\alpha) \,.
\end{eqnarray}
The integration procedure for the high-energy part is adapted to the
problem at hand. To this end, a crude estimate is found for the
dependence of the function $S_{\rm HB}$ defined in Eq.~(\ref{defSHB})
on its arguments.  The considerations leading to this estimate are
analogous to those outlined in Sec.~\ref{InfraRed} for the infrared
part. The result is the approximate expression
\begin{equation}
\label{approxshb}
e^{-y} \,
\exp\left[-\left(\frac{1}{a\,t} - \frac{1}{2}\right)\,(1-r)\,y\right] \,
\end{equation}
for $S_{\rm HB}$.
This leads naturally to the definition
\begin{equation}
\label{DefinitionOfqHB}
q_{\rm HB} = 1 + \left(\frac{1}{a\,t} - \frac{1}{2}\right)\,(1-r)\,,
\end{equation}
so that the (approximate) dependence of $S_{\rm HB}$ on the radial variable
at large $y$ is $\exp\left(-q_{\rm HB}\,y\right)$.
Note that $q_{\rm HB}$ may assume large values ($\gg 1$) as
$t \to 0$; this is unlike the analogous quantity
\[
1 + \left(\frac{c}{a} - \frac{1}{2}\right)\,(1-r)
\]
in the infrared and the middle-energy part, where $|c| =
|\sqrt{1-z^2}| < 1$ because $z \in (0,E_n)$. Having identified the
leading exponential asymptotic behavior of the integrand $S_{\rm HB}$, it is
rather straightforward to evaluate the three-dimensional integral in
Eq.~(\ref{EvaluationOfEHB}) by Gauss-Laguerre and
Gauss-Legendre quadrature~\cite{Je1999}
[the scaling parameter $a$ is defined in Eqs.~(\ref{DefinitionOfA})
and~(\ref{ExpansionOfA})].
The numerical results for the high-energy remainder
function $F_{\rm HB}$ are found in Table~\ref{tableFHKL}.
These results are consistent with the limits
in Eq.~(\ref{limFHB}).

We now turn to a brief discussion of the convergence
acceleration techniques used in the evaluation of 
the function $S_{\rm HB}$ defined in Eq.~(\ref{defSHB}).
The angular momentum decomposition of $S_{\rm HB}$ gives
rise to a sum over the terms $T_{{\rm HB},|\kappa|}$ [see the last
line of Eq.~(\ref{defSHB})], where $|\kappa|$ represents the modulus
of the Dirac angular momentum quantum number of the virtual
intermediate state. In shorthand notation, and suppressing the
arguments, we have
\begin{equation}
\label{defSasymp}
S_{\rm HB} \propto \sum_{|\kappa|=1}^{\infty} T_{{\rm HB},|\kappa|}.
\end{equation}
The radial Green function $G_{\rm B} = G_{\rm B}(r y/a, y/a,z)$ in
coordinate space needs to be evaluated at the radial arguments
$r\,y/a$ and $y/a$ (where $0 < r < 1$), and at the energy argument $z
= E_n - \omega = {\rm i}/2 \, (t^{-1} - t)$ [see Eq.~(\ref{defSHB})].  A
crucial role is played by the ratio $r$ of the two radial arguments.
Indeed, for $|\kappa| \to \infty$, we have [see Eq.~(4.7) in
\cite{Mo1974b}]
\begin{equation}
\label{defTasymp}
T_{{\rm HB},|\kappa|} = \frac{r^{2\,|\kappa|}}{|\kappa|} \,
 \left[{\rm const.} + {\rm O}\!\left(\frac{1}{|\kappa|}\right)\right]\,,
\end{equation}
where ``${\rm const.}$'' is independent of $|\kappa|$ and depends only on
$r$, $y$ and $t$. The series in Eq.\ (\ref{defSasymp}) is slowly
convergent for $r$ close to one, and the region near $r=1$ is known to
be problematic in numerical evaluations. Additionally, note that the
region at $r=1$ is more important at low $Z$ than at high $Z$.  This
is because the function $S_{\rm HB}$, for constant $y$, 
depends on $r$ roughly as $\exp\left[-y\,(1-r)/(a\,t)\right]$
[see Eq.~(\ref{approxshb})], where 
$a = 2\,(Z\alpha)/n + {\rm O}[(Z\alpha)^3]$.  For small
$Z$, the Bohr radius $1/(Z\alpha)$ of the hydrogenlike system is large
compared to high-$Z$ systems, which emphasizes the region near 
$r=1$. 
In this region, the series in (\ref{defSasymp}) is very slowly convergent.
We have found that the convergence of this
series near $r=1$ can be
accelerated very efficiently using the combined nonlinear-condensation
transformation~\cite{JeMoSoWe1999} applied to the
series $\sum_{k=0}^{\infty} t_k$ where $t_k = T_{{\rm HB},k+1}$
[see Eqs.~(\ref{defSasymp}) and~(\ref{defTasymp})]. 

We first transform this series into an
alternating series by a condensation transformation due to Van
Wijngaarden \cite{NP1961,vW1965},
\begin{equation}
\label{vWTrans}
\sum^{\infty}_{k=0} t_k = \sum_{j=0}^{\infty} (-1)^j \, {\bf A}_j \,,
\end{equation}
where 
\begin{equation}
\label{defAj}
{\bf A}_j = \sum^{\infty}_{k=0} 2^k \, t_{2^k\,(j+1)-1}.
\end{equation}
We then accelerate the convergence of the alternating series
$\sum_{j=0}^{\infty} (-1)^j \, {\bf A}_j$ by applying the
nonlinear delta transform $\delta^{(0)}_n (1, {\bf S}_0)$,
which is discussed extensively in~\cite{We1989}.
The explicit formula for this transformation
is given by defining
\begin{equation}
\label{DefSn}
{\bf S}_n = \sum_{j=0}^{n} (-1)^j \, {\bf A}_j 
\end{equation}
as the $n$th partial sum of the Van Wijngaarden transformed input series.
The delta transform reads [see Eq.~(8.4-4) of~\cite{We1989}],
\begin{equation}
\label{dWenTr}
{\delta}_n^{(0)} (1, {\bf S}_0) \; = \;
\frac {\displaystyle
\sum_{j=0}^{n} \; (- 1)^{j} \; {{n} \choose {j}} \;
\frac {(1 + j)_{n-1}} {(1 + n)_{n-1}} \;
\frac {{\bf S}_{j}} {{\bf B}_{j + 1}} }
{\displaystyle
\sum_{j=0}^{n} \; (- 1)^{j} \; {{n} \choose {j}} \;
\frac {(1 + j)_{n-1}} {(1 + n)_{n-1}} \;
\frac {1} {{\bf B}_{j + 1}} } \, ,
\end{equation}
where 
\begin{equation} 
\label{DefBn}
{\bf B}_j = (-1)^j \, {\bf A}_j \,.
\end{equation}
The convergence acceleration proceeds by calculating a sequence of 
transforms ${\delta}_n^{(0)}$ in increasing transformation order
$n$. It is observed that the transforms
converge much faster than the partial sums ${\bf S}_n$ defined in
Eq.~(\ref{DefSn}).
The upper index zero in Eq.~(\ref{dWenTr}) indicates that the
transformation is started with the first term ${\bf A}_0$.

The combined transformation (combination of
the condensation transformation and the Weniger transformation) was
found to be applicable to a wide range of slowly convergent monotone
series (series whose terms have the same sign), and many examples for
its application were given in Ref.~\cite{JeMoSoWe1999}. For the
numerical treatment of radiative corrections in low-$Z$ systems, the
transformation has the advantage of removing the principal 
numerical difficulties
associated with the slow convergence of angular momentum
decompositions of the propagators near their singularity for equal
radial arguments.

In a typical case, sufficient precision ($10^{-11}$) in the
convergence of the sum in Eq.\ (\ref{defSasymp}) is reached in a
transformation order $n < 100$ for the nonlinear transformation
$\delta^{(0)}_n (1, {\bf S}_0)$, a region in which the nonlinear
sequence transformation $\delta$ is numerically stable.  
Although the delta transformation exhibits considerable
numerical stability in higher transformation 
orders~\cite{We1989,JeMoSoWe1999}, inevitable round-off errors
start to accumulate significantly in an excessively high
transformation order of $n \approx 500$ in a typical
case~\cite{Je1999}, and this situation is avoided in the current
evaluation because the transforms exhibit apparent convergence
to the required accuracy before numerical round-off errors
accumulate. Note that
evaluation of the condensed series ${\bf A}_j$ in Eq.\ (\ref{defAj})
entails sampling of terms $T_{{\rm HB},|\kappa|}$ for rather large
$|\kappa|$, while eliminating the necessity of evaluating {\em all}
terms $T_{{\rm HB},|\kappa|}$ up to the maximum index. The highest
angular momentum $|\kappa|$ encountered in the present calculation is
in excess of $4\,000\,000$. However, even in extreme cases less than
$3\,000$ evaluations of particular terms of the original series are
required.  The computer time for the evaluation of the slowly
convergent angular momentum expansion near the singularity is reduced
by roughly three orders of magnitude by the use of the
convergence acceleration methods. 

In certain parameter 
regions (e.g.~for large energy of the virtual photon), 
a number of terms of the input series $t_k$ have to
be skipped before the convergence acceleration algorithm defined in
Eqs.~(\ref{vWTrans})--(\ref{DefBn}) can be applied
(in order to avoid transient behavior of the first few terms 
in the sum over $\kappa$). 
In this case, the input data for the combined
nonlinear-condensation transformation are
the terms $t_k = T_{{\rm HB},k+1+\kappa_s}$,
where $\kappa_s$ denotes the number of terms which are
directly summed before the transformation is applied.
These issues and further details regarding
the application of the convergence acceleration method
to QED calculations can be found in Appendix H.2 of~\cite{Je1999}.

\typeout{Subsection 4C}
%
%
\subsection{Results for the High-Energy Part}

The limit of the function $F_{\rm H}$ as $Z\alpha\to0$
can be derived easily from Eqs.~(\ref{limFHAgen}),
(\ref{limFHBgen}) as a function of the bound state quantum
numbers. For $F_{\rm H}$ the limit is
\begin{eqnarray}
\lefteqn{F_{\rm H}(nl_j,Z\alpha) =}
\nonumber\\[2ex] 
&=& \left(\frac{11}{10} - \ln2\right) \, \frac{1}{n}
+ \left(\frac{16}{15} - 2\,\ln2\right) \, \frac{1}{2\,l + 1} 
\nonumber\\[1ex]
& & + \left(-\frac{3}{2}\,\ln2 + \frac{5}{4}\right)\, 
\frac{1}{\kappa\,(2\,l+1)} +
\left(\frac{3}{2}\,\ln2 - \frac{9}{4}\right)\,\frac{1}{|\kappa|} 
\nonumber \\[1ex]
& & + 
\left(\frac{17}{18} - \frac{4}{3}\,\ln2\right) \delta_{l,0} +
\left(\frac{5}{6} - \ln2\right) \frac{n-2\,l-1}{n\,(2\,l+1)} 
\nonumber \\[1ex]
& & + 
{\rm O}(Z\alpha)\,.
\end{eqnarray}
For the atomic states investigated here, 
this expression yields the numerical values 
\begin{eqnarray}
\label{limFH}
F_{\rm H}(1{\rm S}_{1/2},Z\alpha) &=&
- 1.313~085 + {\rm O}(Z\alpha)\,, \nonumber\\[2ex]
F_{\rm H}(2{\rm S}_{1/2},Z\alpha) &=&
- 1.446~418 + {\rm O}(Z\alpha)\,, \nonumber\\[2ex]
F_{\rm H}(2{\rm P}_{1/2},Z\alpha) &=&
- 1.066~667 + {\rm O}(Z\alpha)\,, \nonumber\\[2ex]
F_{\rm H}(2{\rm P}_{3/2},Z\alpha) &=&
- 0.566~667 + {\rm O}(Z\alpha)\,.
\end{eqnarray}
Numerical results for the high-energy part
\begin{equation}
F_{\rm H}(nl_j,Z\alpha) = 
F_{\rm HA}(nl_j,Z\alpha) + 
F_{\rm HB}(nl_j,Z\alpha)
\end{equation}
are also summarized in Table~\ref{tableFHKL}. 
Note the apparent consistency
of the numerical results in Table~\ref{tableFHKL} with their analytically
obtained low-$Z$ limits in Eq.~(\ref{limFH}).


\widetext

%
%
\begin{table}[htb]
\begin{center}
\begin{minipage}{17cm}
\begin{center}
\caption{\label{tableFKL} Numerical results for
the scaled self-energy function
$F$ and the self-energy remainder function $G_{\rm SE}$.}
\begin{tabular}{lr@{.}lr@{.}lr@{.}lr@{.}l}
$Z$ & 
\multicolumn{2}{c}{\rule[-3mm]{0mm}{8mm} 
  $F(1{\rm S}_{1/2},Z\alpha)$} &
\multicolumn{2}{c}{\rule[-3mm]{0mm}{8mm} 
  $F(2{\rm S}_{1/2},Z\alpha)$} &
\multicolumn{2}{c}{\rule[-3mm]{0mm}{8mm} 
  $F(2{\rm P}_{1/2},Z\alpha)$} &
\multicolumn{2}{c}{\rule[-3mm]{0mm}{8mm} 
  $F(2{\rm P}_{3/2},Z\alpha)$} \\
\hline
$1$ &
 $10$ & $316~793~659(1)$ &
 $10$ & $546~825~185(5)$ &
 $-0$ & $126~396~37(1)$ &
 $ 0$ & $123~498~56(1)$ \\
$2$ &
  $8$ & $528~325~061(1)$ &
  $8$ & $758~870~25(1)$ &
 $-0$ & $125~816~16(1)$ &
 $ 0$ & $123~835~55(1)$ \\
$3$ &
  $7$ & $504~503~432(1)$ &
  $7$ & $735~777~20(1)$ &
 $-0$ & $124~992~24(1)$ &
 $ 0$ & $124~317~10(1)$ \\
$4$ &
  $6$ & $792~824~089(1)$ &
  $7$ & $025~002~41(1)$ &
 $-0$ & $123~968~79(1)$ &
 $ 0$ & $124~918~48(1)$ \\
$5$ &
  $6$ & $251~627~086(1)$ &
  $6$ & $484~860~42(1)$ &
 $-0$ & $122~774~94(1)$ &
 $ 0$ & $125~623~30(1)$ \\
\hline
\hline
$Z$ & 
\multicolumn{2}{c}{\rule[-3mm]{0mm}{8mm} 
  $G_{\rm SE}(1{\rm S}_{1/2},Z\alpha)$} &
\multicolumn{2}{c}{\rule[-3mm]{0mm}{8mm} 
  $G_{\rm SE}(2{\rm S}_{1/2},Z\alpha)$} &
\multicolumn{2}{c}{\rule[-3mm]{0mm}{8mm} 
  $G_{\rm SE}(2{\rm P}_{1/2},Z\alpha)$} &
\multicolumn{2}{c}{\rule[-3mm]{0mm}{8mm} 
  $G_{\rm SE}(2{\rm P}_{3/2},Z\alpha)$} \\
\hline
$1$ &
 $-30$ & $290~24(2)$ &
 $-31$ & $185~15(9)$ &
 $ -0$ & $973~5(2)$ &
 $ -0$ & $486~5(2)$ \\
$2$ &
 $-29$ & $770~967(5)$ &
 $-30$ & $644~66(5)$ &
 $ -0$ & $949~40(5)$ &
 $ -0$ & $470~94(5)$ \\
$3$ &
 $-29$ & $299~169(2)$ &
 $-30$ & $151~93(2)$ &
 $ -0$ & $926~37(2)$ &
 $ -0$ & $456~65(2)$ \\
$4$ &
 $-28$ & $859~223(1)$ &
 $-29$ & $691~27(1)$ &
 $ -0$ & $904~12(1)$ &
 $ -0$ & $443~13(1)$ \\
$5$ &
 $-28$ & $443~372~3(8)$\tablenote{The result for this entry
given in~\protect\cite{JeMoSo1999} contains a typographical error.} &
 $-29$ & $255~033(8)$ &
 $ -0$ & $882~478(8)$ &
 $ -0$ & $430~244(8)$ \\
\end{tabular}
\end{center}
\end{minipage}
\end{center}
\end{table}

\narrowtext

\typeout{Section 5}
%
%
\section{Comparison to Analytic Calculations}
\label{ComparisonAnalytic}

The numerical results for the scaled self-energy function 
$F(nl_j,Z\alpha)$ defined in Eq.~(\ref{ESEasF}) are given in
Table~\ref{tableFKL}, together with the results for the nonperturbative
self-energy remainder function $G_{\rm SE}(nl_j,Z\alpha)$,
which is implicitly defined in Eq.~(\ref{defFLO}). 
Results are provided for K- and L-shell states. 
The results here at $Z=5$ are consistent with and much more precise 
than the best previous calculation \cite{Mo1992}.
The numerical
results for the self-energy remainder $G_{\rm SE}$ are obtained by
subtracting the analytic
lower-order terms listed in Eq.~(\ref{defFLO}) from the
complete numerical result for the scaled self-energy function 
$F(nl_j,Z\alpha)$. No additional fitting is performed. 

Analytic and numerical results at low
$Z$ can be compared by considering the self-energy remainder
function $G_{\rm SE}$. Note that an inconsistency in any
of the analytically obtained lower-order terms would be likely
to manifest itself in a grossly inconsistent dependence of
$G_{\rm SE}(nl_j,Z\alpha)$ on its argument $Z\alpha$;
this is not observed.
For S states, the following analytic
model for $G_{\rm SE}$ is commonly assumed,
which is motivated in part
by a renormalization-group analysis~\cite{MaSt2000} and is
constructed
in analogy with the pattern of the analytic coefficients $A_{ij}$
in Eq.~(\ref{defFLO}) and~(\ref{defFLOnS})
\begin{eqnarray}
\lefteqn{G_{\rm SE}(n{\rm S}_{1/2},Z\alpha) =
A_{60}(n{\rm S}_{1/2})} \nonumber\\[1ex]
& & \quad + (Z\alpha) \,
\left[A_{71}(n{\rm S}_{1/2}) \, \ln(Z\alpha)^{-2} 
  + A_{70}(n{\rm S}_{1/2}) \right] 
\nonumber\\[1ex]
& & \quad + (Z\alpha)^2 \,
\left[
A_{83}(n{\rm S}_{1/2}) \, \ln^3(Z\alpha)^{-2} \right.
\nonumber\\[1ex]
& & \qquad\qquad \left. + 
A_{82}(n{\rm S}_{1/2}) \, \ln^2(Z\alpha)^{-2} \right.
\nonumber\\[1ex]
& & \qquad\qquad \left. 
+ A_{81}(n{\rm S}_{1/2}) \, \ln(Z\alpha)^{-2} 
+ A_{80}(n{\rm S}_{1/2}) \right]\,.
\end{eqnarray}
The (probably nonvanishing) $A_{83}$ coefficient, which 
introduces a triple logarithmic singularity at $Z\alpha=0$, 
hinders an accurate comparison of numerical and 
analytic data for $G_{\rm SE}$. 
A somewhat less singular behavior is expected of the 
difference
\begin{equation}
\label{DefDeltaGSE}
\Delta G_{\rm SE}(Z\alpha) = 
G_{\rm SE}(2{\rm S}_{1/2},Z\alpha) -
G_{\rm SE}(1{\rm S}_{1/2},Z\alpha)\,,
\end{equation}
because the leading logarithmic coefficients in any given order
of $Z\alpha$ are generally assumed to be equal for all
S states, which would mean in particular
\begin{eqnarray}
A_{71}(1{\rm S}_{1/2}) &=& A_{71}(2{\rm S}_{1/2})
\quad \mbox{and}
\nonumber\\[2ex]
A_{83}(1{\rm S}_{1/2}) &=& A_{83}(2{\rm S}_{1/2})\,.
\end{eqnarray}
Now we define $\Delta A_{kl}$ as the difference of the values of the 
analytic coefficients for the two lowest S states:
\begin{equation}
\label{DefDeltaA}
\Delta A_{kl} = 
A_{kl}(2{\rm S}_{1/2}) -
A_{kl}(1{\rm S}_{1/2})\,.
\end{equation}
The function $\Delta G_{\rm SE}$ defined in
Eq.~(\ref{DefDeltaGSE}) can be assumed to have the following
semi-analytic expansion about $Z \alpha = 0$:
\begin{eqnarray}
\label{ExpansionDeltaGSE}
\lefteqn{\Delta G_{\rm SE}(Z\alpha) \; = \;
\Delta A_{60} + (Z\alpha) \, \Delta A_{70}}
\nonumber\\[1ex]
& & \quad + (Z\alpha)^2 \,
\left[
\Delta A_{82} \, \ln^2(Z\alpha)^{-2} \right.
\nonumber\\[1ex]
& & \qquad \left. 
+ \Delta A_{81} \, \ln(Z\alpha)^{-2} 
+ \Delta A_{80} 
+ {\rm o}(Z\alpha) \right]\,.
\end{eqnarray}
In order to detect possible inconsistencies in the 
numerical and analytic data for $G_{\rm SE}$,  
we difference the data for $\Delta G_{\rm SE}$,
i.e., we consider the following finite difference approximation
to the derivative of the function $\Delta G_{\rm SE}$:
\begin{equation}
\label{DefgZ}
g(Z) = \Delta G_{\rm SE}\bbox{(} (Z+1)\,\alpha\bbox{)} -
\Delta G_{\rm SE}\bbox{(}Z\alpha\bbox{)}\,.
\end{equation}
We denote the analytic
and numerical limits of $\Delta G_{\rm SE}(Z\alpha)$ as
$Z\alpha \to 0$ as $\Delta A^{\rm (an)}_{60}$ and $\Delta A^{\rm (nu)}_{60}$,
respectively, 
and leave open the possibility of 
an inconsistency between numerical and analytic data by keeping
$\Delta A^{\rm (nu)}_{60}$ and $\Delta A^{\rm (an)}_{60}$
as distinct variables.
In order to illustrate how a discrepancy
could be detected by investigating the function
$g(Z)$, we consider special cases
of the function $\Delta G_{\rm SE}(Z\alpha)$ and $g(Z)$.
We have for $Z=0$, which is determined exclusively by analytic 
results,
\begin{equation}
\Delta G_{\rm SE}(0) = \Delta A^{\rm (an)}_{60}\,,
\end{equation}
whereas for $Z=1$, which is determined by numerical data,
\begin{equation}
\Delta G_{\rm SE}(\alpha) = \Delta A^{\rm (nu)}_{60} +
  \alpha\,\left[ \Delta A_{70} + {\rm o}(\alpha) \right]\,,
\end{equation}
and for $Z=2$,
\begin{equation}
\Delta G_{\rm SE}(2\alpha) = \Delta A^{\rm (nu)}_{60} +
  \alpha\,\left[ 2\,\Delta A_{70} + {\rm o}(\alpha) \right]\,,
\end{equation}
etc. Hence for $Z=0$, we have 
\begin{eqnarray}
\label{kink}
\lefteqn{g(0) = \Delta G_{\rm SE}(\alpha) - \Delta G_{\rm SE}(0)} 
\nonumber\\[2ex]
&=& \Delta A^{\rm (nu)}_{60} - \Delta A^{\rm (an)}_{60} +
  \alpha\,\left[ \Delta A_{70} + {\rm o}(Z\alpha) \right]\,.
\end{eqnarray}
For $Z=1$, the value of $g$ is determined solely by numerical data,
\begin{eqnarray}
g(1) &=& \Delta G_{\rm SE}(2\alpha) - \Delta G_{\rm SE}(\alpha) 
\nonumber\\[2ex]
&=& \alpha\,\left[ \Delta A_{70} + {\rm o}(Z\alpha) \right]\,,
\end{eqnarray}
and for $Z=2$, we have
\begin{eqnarray}
\label{g2}
g(2) &=& \Delta G_{\rm SE}(3\alpha) - \Delta G_{\rm SE}(2\alpha)         
\nonumber\\[2ex]
&=& \alpha\,\left[ \Delta A_{70} + {\rm o}(Z\alpha) \right]\,.
\end{eqnarray}
Analogous equations hold for $Z > 2$. The analytic and the
numerical data
from Table~\ref{tableFKL} lead to the five values
$g(0)$, $g(1)$, $g(2)$, $g(3)$, and $g(4)$. 
A plot of the
function $g(Z)$ serves two purposes: First, the values
$g(1),\dots,g(4)$ should exhibit apparent convergence to
some limiting value 
$\alpha\,\Delta A_{70}$ as $Z \to 0$, and this can
be verified by inspection of the plot. Secondly,
a discrepancy between the analytic and numerical approaches
would result in a nonvanishing value for 
$\Delta A^{\rm (nu)}_{60} - \Delta A^{\rm (an)}_{60}$ which 
would appear as an inconsistency between the trend in the values of
$g(1),\dots,$ and $g(4)$ and the value of $g(0)$ [see Eq.~(\ref{kink})].

%
%
\begin{figure}[htb]
\begin{center}
\begin{minipage}{8.6cm}
\centerline{\mbox{\epsfysize=7.0cm\epsffile{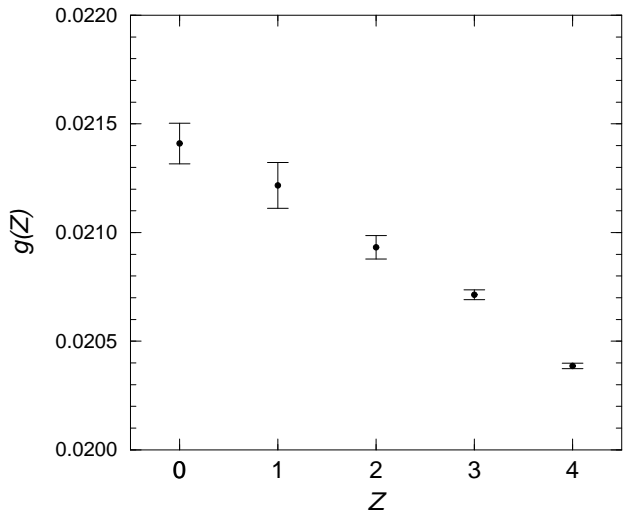}}}
\caption{\label{Plot2S1S} Plot of the function $g(Z)$ defined in
Eq.~(\protect\ref{DefgZ}) in the
region of low nuclear charge.
For the evaluation of the 
data point at $Z = 0$, a value of $A_{60}(1{\rm S}_{1/2}) =
-30.924\,15(1)$ is employed~\protect\cite{Pa1993,JeMoSo1999,Pa1998priv}.}
\end{minipage}
\end{center}
\end{figure}

Among the separate evaluations of $A_{60}$ for the ground state, the
result in~\cite{Pa1993} has the smallest quoted uncertainty.  In
Fig.~\ref{Plot2S1S} we display a plot of $g(Z)$ for low nuclear charge
$Z$. A value of $A_{60}(1{\rm S}_{1/2}) = 
A^{\rm (an)}_{60}(1{\rm S}_{1/2}) = 
-30.92415(1)$~\cite{Pa1993,JeMoSo1999,Pa1998priv}
is used in Fig.~\ref{Plot2S1S}.
The results indicate very good agreement between the numerical and
analytic approaches to the Lamb shift in the low-$Z$ region up to the
level of a few Hz in frequency units for the low-lying
atomic states (where $n$ is the principal quantum number).
The error bars represent the numerical uncertainty of the 
values in Table~\ref{tableFKL}, which correspond to an 
uncertainty on the level of $1.0 \times Z^4 \, {\rm Hz}$ in
frequency units.

Analytic work on the correction $A_{60}$ has
extended over three decades~\cite{ErYe1965a,ErYe1965b,Er1971,Sa1981,Pa1993}.  
The complication arises that although the calculations are in general
analytic, some remaining one-dimensional integrations could not be
evaluated analytically because of the nature of the integrands
[see e.g.~Eq.~(6.96) in~\cite{Pa1993}]. Therefore a
step-by-step comparison of the analytic calculations is difficult.  An
additional difficulty is the isolation of those analytic terms which
contribute in a given order in $Z\alpha$, i.e.,~the isolation of only
those terms which contribute to $A_{60}$. The apparent consistency 
of the numerical and analytic data in Fig.~\ref{Plot2S1S} represents
an independent consistency check on the rather
involved analytic calculations.
 
%
%
\begin{figure}[htb]
\begin{center}
\begin{minipage}{8.6cm}
\centerline{\mbox{\epsfysize=7.0cm\epsffile{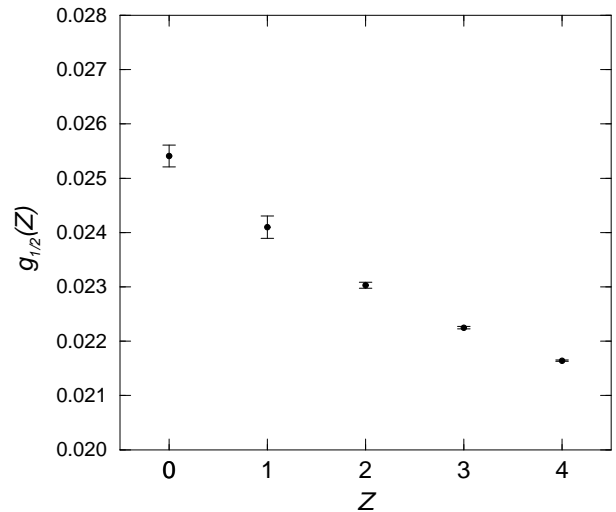}}}
\caption{\label{Plot2P12} Comparison of numerical data
and analytically evaluated higher-order binding corrections
for the $2{\rm P}_{1/2}$ state.
We plot the function $g_{1/2}(Z)$ defined
in Eq.~(\ref{Defgj}) in the region of low $Z$. The numerical data
obtained in the current investigation appear to be consistent with the
analytic result of
$A_{60}(2{\rm P}_{1/2}) = -0.998\,91(1)$
obtained in~\protect\cite{JePa1996}.}

\end{minipage}
\end{center}
\end{figure}

Our numerical results are not inconsistent with the
analytic result \cite{Ka1997} for a higher-order logarithm,
\begin{equation}
A_{71} = \pi\,\left(\frac{139}{64}-\ln2\right) = 4.65,
\end{equation}
although they do not necessarily confirm it. As
in~\cite{JeMoSo1999}, we obtain as an estimate $A_{71} = 5.5(1.0)$
(from the fit to the numerical data for both S states). Logarithmic
terms corresponding to the (probably) nonvanishing $A_{83}$ coefficient
should be taken into account for a consistent fit of the
corrections to $G_{\rm SE}$. These highly singular terms are difficult
to handle with a numerical fitting procedure. The terms 
$A_{83}$, $A_{82}$ and $A_{81}$ furnish
three more free parameters for the numerical fit, where only five data
points are available (in addition to the quantities $A_{60}$, $A_{71}$
and $A_{70}$, which may also be regarded as free parameters for the
fitting procedure).  The determination of $A_{60}$ by a fit from the
numerical data is much more stable than the determination of
the logarithmic correction $A_{71}$. We briefly note that our
all-order evaluation essentially eliminates the uncertainty due to the
unknown higher-order analytic terms. 
Also, it is interesting to note that the same numerical 
methods are employed for both the S and P states in our all-order
(in $Z\alpha$) calculation, whereas the analytic treatment
of S and P states differs~\cite{Pa1993,JePa1996}.

\begin{figure}[htb]
\begin{center}
\begin{minipage}{8.6cm}
\centerline{\mbox{\epsfysize=7.0cm\epsffile{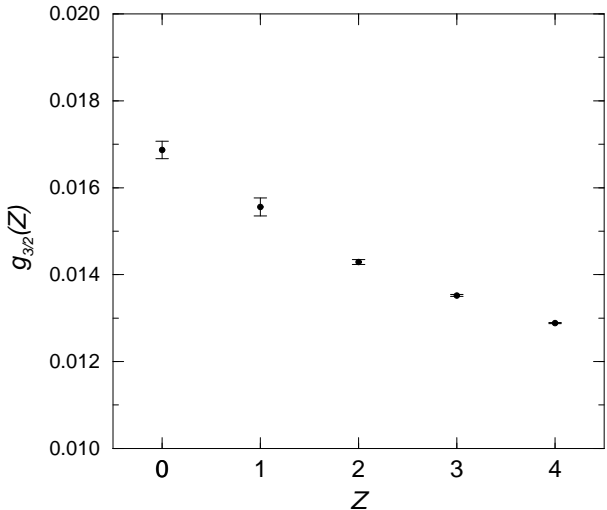}}}
\caption{\label{Plot2P32} For the $2{\rm P}_{3/2}$ state,
we plot the function $g_{3/2}(Z)$ defined
in Eq.~(\ref{Defgj}) in the region of low $Z$. The numerical data
obtained in the current investigation appear to be consistent with
the analytic result of 
$A_{60}(2{\rm P}_{3/2}) = -0.503\,37(1)$
from~\protect\cite{JePa1996}.}
\end{minipage}
\end{center}
\end{figure}

The comparison of numerical and analytic results is much 
less problematic for P states, because the function
$G_{\rm SE}$ is less singular [see Eqs.~(\ref{defFLOnPj})
and~(\ref{A60nP})].
For the 2P states, 
we observe that the function $G_{\rm SE}(2{\rm P}_j,Z\alpha)$ 
has the same semi-analytic expansion about $Z\alpha=0$ as the 
function $\Delta G_{\rm SE}(Z\alpha)$ defined for S states in
Eq.~(\ref{DefDeltaGSE}). We have
\begin{eqnarray}
\lefteqn{G_{\rm SE}(2{\rm P}_j,Z\alpha) \; = \;
A_{60}(2{\rm P}_j) + (Z\alpha) \, A_{70}(2{\rm P}_j)}
\nonumber\\[1ex]
& & \quad + (Z\alpha)^2 \,
\left[
A_{82}(2{\rm P}_j) \, \ln^2(Z\alpha)^{-2} \right.
\nonumber\\[1ex]
& & \qquad \left. 
+ A_{81}(2{\rm P}_j) \, \ln(Z\alpha)^{-2} 
+ A_{80}(2{\rm P}_j) 
+ {\rm o}(Z\alpha) \right]\,.
\end{eqnarray}
Hence, we plot the function
\begin{equation}
\label{Defgj}
g_j(Z) =
G_{\rm SE}\bbox{(}2{\rm P}_j,(Z+1)\,\alpha\bbox{)} -
G_{\rm SE}\bbox{(}2{\rm P}_j,Z\alpha\bbox{)}
\end{equation}
for $j=1/2$ and $j=3/2$ in the region of low $Z$, with the notion that
an inconsistent analytic result for $A_{60}(2{\rm P}_j)$ would lead to
irregularity at $Z=0$, in analogy with the
S states. The numerical data
shown in Figs.~\ref{Plot2P12}, 
and~\ref{Plot2P32} appear to be
consistent with the analytic results of
\begin{eqnarray}
A_{60}(2{\rm P}_{1/2}) &=& -0.998\,91(1) \quad \mbox{and}
\nonumber\\[2ex]
A_{60}(2{\rm P}_{3/2}) &=& -0.503\,37(1)
\end{eqnarray}
obtained in~\cite{JePa1996}. In this context it may be interesting to
note that analytic results obtained in~\cite{JePa1996,JeSoMo1997}
for the higher-order binding corrections to 2P, 3P, and 4P states have
recently been confirmed indirectly~\cite{Pa1999}.
Finally, although it may be 
possible to obtain more accurate estimates of some
higher-order analytic corrections, notably the
$A_{70}$ coefficient for P states and $\Delta A_{70}$
for the two lowest-lying S states, 
we have not made such an analysis in the current work;
we have restricted the discussion to a check of the 
consistency with the available results for $A_{60}$.

%
%


\typeout{Section 6}
%
%
\section{Conclusion}
\label{Conclusion}

There has recently been a rather broad interest in the
numerical calculation of relativistic, QED self energy, and
two-body corrections at low $Z$ and the comparison
of analytic and numerical results~\cite{ArShYe1995pra,ArShYe1995,
ShEtAl1998jpb,ShEtAl1998,PaGr1995,
Ye1998,BeSo1988,ScGrSo1993,Ka1993log,
MaSt2000,GoLaNePlSo1999,MaSa1998b,Ye2000}.
Traditionally, the self-energy correction for hydrogenlike systems
has posed a computational challenge.
Here we have described a nonperturbative evaluation of the one-photon
self-energy correction in hydrogenlike ions with low nuclear charge
numbers $Z=1$ to $5$. The general outline of our approach is discussed
in Sec.~\ref{MethodOfEvaluation}. In Sec.~\ref{LowEnergyPart}, the
numerical evaluation of the low-energy part (generated by virtual
photons of low energy) is described. In Sec.~\ref{HighEnergyPart}, we
discuss the numerical evaluation of the high-energy part, which is
generated by high-energy virtual photons and contains the formally
infinite contributions, which are removed by the
renormalization. Sec.~\ref{HighEnergyPart} also contains a brief
discussion of the convergence acceleration methods as employed in the
current evaluation. We discuss in Sec.~\ref{ComparisonAnalytic} the
comparison of analytic and numerical data for K- and L-shell states in
the region of low $Z$. The main results of this paper are contained in
Table~\ref{tableFKL}: numerical data, nonperturbative in $Z\alpha$,
for the scaled self-energy function $F$ and the self-energy remainder
function $G_{\rm SE}$ for K- and L-shell states at low nuclear charge.
The numerical accuracy of our data is 1~Hz or better in frequency
units for 1S, 2S and both 2P states in atomic hydrogen.

The comparison of analytic and numerical results to the level of
accuracy of the numerical data, which is discussed in
Sec.~\ref{ComparisonAnalytic}, indicates that there is 
very good agreement for the K- and L-shell states.
The analytic and numerical data are shown in
Figs.~\ref{Plot2S1S}, \ref{Plot2P12}, and~\ref{Plot2P32}.
Our all-order
evaluation eliminates any uncertainty due to the unknown
higher-order analytic terms; the current numerical uncertainty
in the self energy is at the level of 1~Hz for atomic
hydrogen.

\section*{Acknowledgments}

U. D. J. thanks the National Institute of Standards and Technology for
kind hospitality during a number of extended research appointments. 
He would also like to
acknowledge support from the Deutscher Akademischer Austauschdienst
(DAAD).
The authors would like to acknowledge helpful discussions
with K. Pachucki, S. Karshenboim and J. Sims. P.~J.~M.~acknowledges the
Alexander von Humboldt Foundation for continued support. The authors
wish to acknowledge support from BMBF, DFG and from GSI.

\end{multicols}

\begin{thebibliography}{10}
\bibitem{Disclaimer}
Certain commercial equipment, instruments, or materials are identified in this
  paper to foster understanding. Such identification does not imply
  recommendation or endorsement by the National Institute of Standards and
  Technology, nor does it imply that the materials or equipment identified are
  necessarily the best available for the purpose.
\end{thebibliography}
\end{document}